\RequirePackage{ifpdf}
\ifpdf % We are running pdfTeX in pdf mode
\documentclass[pdftex]{sigma}
\else
\documentclass{sigma}
\fi

\usepackage{amsxtra}
\usepackage{eucal}

\def\({\left(}
\def\){\right)}
\newcommand{\Rdr}{R_\mathrm{dress}}

\newcommand{\phib}{\mbox{\boldmath$\phi$}}
\newcommand{\betab}{\mbox{\boldmath$\beta$}}
\newcommand{\gammab}{\mbox{\boldmath$\gamma$}}

\newcommand{\rhob}{\rho^\mathrm{sc}}
\newcommand{\omegab}{\omega^\mathrm{sc}}
\newcommand{\Tb}{T^\mathrm{sc}}
\newcommand{\Qb}{Q^\mathrm{sc}}

\newcommand{\taub}{\mbox{\boldmath$\tau$}}

\newcommand{\cb}{\mathbf{c}}
\newcommand{\bb}{\mathbf{b}}

\newcommand{\tb}{\mathbf{t}}

\newcommand{\nn}{\nonumber}

\newcommand{\R}{{\mathbb R}}

\newcommand{\slth}{\widehat{\mathfrak{sl}}_2}
\newcommand{\res}{{\rm res}}

\newcommand{\Tr}{{\rm Tr}}

\newcommand{\isoto}[1][]%
{{\mathop{\buildrel{\sim}\over\longrightarrow}\limits_{#1}}}

\def\[{\left[}
\def\]{\right]}
\newcommand{\la}{\lambda}

\newcommand{\al}{\alpha}

\newcommand{\z}{\zeta}

\newcommand{\om}{{\omega}}

\numberwithin{equation}{section}

\newcommand{\bc}{\mathbf{c}}

\begin{document}

\allowdisplaybreaks

\renewcommand{\thefootnote}{$\star$}

\renewcommand{\PaperNumber}{007}

\FirstPageHeading

\ShortArticleName{Fermionic Basis in CFT and TBA for Excited States}

\ArticleName{Fermionic Basis in Conformal Field Theory and\\
Thermodynamic Bethe Ansatz for Excited States\footnote{This paper is a
contribution to the Special Issue ``Relationship of Orthogonal Polynomials and Special Functions with Quantum Groups and Integrable Systems''. The
full collection is available at
\href{http://www.emis.de/journals/SIGMA/OPSF.html}{http://www.emis.de/journals/SIGMA/OPSF.html}}}

\Author{Hermann BOOS~$^{\dag\ddag}$}

\AuthorNameForHeading{H. Boos}

\Address{$^\dag$~Fachbereich C -- Physik, Bergische Universit\"at
Wuppertal, 42097 Wuppertal, Germany}
\EmailD{\href{mailto:boos@physik.uni-wuppertal.de}{boos@physik.uni-wuppertal.de}}

\Address{$^\ddag$~Skobeltsyn Institute of Nuclear Physics, Moscow State
University, 119991 Moscow, Russia}

\ArticleDates{Received October 07, 2010, in f\/inal form January 05, 2011;  Published online January 13, 2011}

\Abstract{We generalize the results of [{\em Comm. Math. Phys.}
\textbf{299} (2010), 825--866] (hidden Grassmann structure IV) to
the case of excited states of the transfer matrix of the six-vertex
model acting in the so-called Matsubara direction. We establish an
equivalence between a scaling limit of the partition function of the
six-vertex model on a cylinder with quasi-local operators inserted
and special boundary conditions, corresponding to particle-hole
excitations, on the one hand, and certain three-point correlation
functions of conformal f\/ield theory (CFT) on the other hand. As in
hidden Grassmann structure~IV, the fermionic basis developed in
previous papers and its conformal limit are used for a description
of the quasi-local operators. In paper IV we claimed that in the
conformal limit the fermionic creation operators  generate a basis
equivalent to the basis of the descendant states in the conformal
f\/ield theory modulo integrals of motion suggested by
A.~Zamolodchikov (1987). Here we argue that, in order to completely
determine the transformation between the above fermionic basis and
the basis of descendants in the CFT, we need to involve excitations.
On the side of the lattice model we use the excited-state TBA
approach. We consider in detail the case of the descendant at level
8.}


\Classification{82B20; 82B21; 82B23; 81T40; 81Q80}

\renewcommand{\thefootnote}{\arabic{footnote}}
\setcounter{footnote}{0}

\section{Introduction}

Much progress was made in the understanding of the connection
between the one-dimensional XXZ spin chain and two-dimensional
quantum f\/ield theories (QFT). Many dif\/ferent aspects of this
connection were studied in the literature. In the present paper we
will touch only some particular aspect related to the hidden
fermionic structure of the XXZ model and the corresponding continuum
model -- conformal f\/ield theory (CFT),
\cite{HGSI,HGSII,HGSIII,HGSIV}.

The continuum model can be studied through the scaling limit of the
six-vertex model compactif\/ied on an inf\/inite cylinder of radius $R$.
The corresponding direction around the cylinder is sometimes called
Matsubara direction. It is well known that the scaling limit of the
homogeneous critical six-vertex model is related to CFT, while
introducing inhomogeneities in a~special way leads to the
sine-Gordon model (sG) which has a mass gap, \cite{Destri-deVega}.
An important step forward was done by Bazhanov, Lukyanov and
Zamolodchikov in papers \cite{BLZI,BLZII,BLZIII}. They obtained an
integrable structure of CFT by constructing a monodromy matrix with
the quantum space related to the chiral bosonic f\/ield and the
Heisenberg algebra. On the other hand this monodromy matrix
satisf\/ies the Yang--Baxter equation with the $R$-matrix of the
six-vertex model. The corresponding transfer matrix fulf\/ills
Baxter's $TQ$-relation \cite{baxter} and generates the integrals of
motion. These integrals of motion are special combinations of the
Virasoro generators. Originally they were introduced by Alexander
Zamolodchikov in \cite{Zam}. Zamolodchikov observed that the
integrals of motion generate that part of the Virasoro algebra that
survives under the integrable $\Phi_{1,3}$-perturbation of the CFT.

\looseness=1
In order to state the full equivalence of the
six-vertex model in the scaling limit and the CFT one needs to
compare all possible correlation functions. This problem is far from
being solved completely. We believe that it can be helpful to use a
hidden fermionic structure of the spin-$\frac12$ XXZ chain,
\cite{HGSI,HGSII}. The key idea is to consider a fermionic
basis generated by means of certain creation operators $\tb ^*$,
$\bb ^*$, $\cb ^*$. These operators act on a space of quasi-local
operators $\mathcal{W}^{(\al)}$. Any operator in this space can be
represented in the form $q^{2\al S(0)}\mathcal{O}$ with some
operator $\mathcal{O}$ which acts on a chain segment of arbitrary
f\/inite length. The operator~$q^{2\al S(0)}$ was called ``primary
f\/ield'' because it fulf\/ills some properties similar to the
properties of certain primary f\/ield in CFT. The fermionic basis is
constructed inductively. Its completeness was shown in
\cite{complet}. An important theorem proved by Jimbo, Miwa and
Smirnov in the paper \cite{HGSIII} allows one to reduce any
correlation function expressed through the fermionic basis to
determinants. An interesting feature of the above construction is
that it is algebraic in the sense that the fermionic operators are
constructed by means of the representation theory of the quantum
group $U_q(\slth)$. They are independent of any physical data like
the magnetic f\/ield, the temperature, the boundary conditions etc.
For example, the temperature can be incorporated via the
Suzuki--Trotter formalism \cite{SuzukiTrotter} by taking
inhomogeneity parameters in the Matsubara direction in a special way
and then performing the so-called Trotter limit, when the number of
sites in the Matsubara direction $\mathbf{n}\rightarrow\infty$. One
can even keep the number $\mathbf{n}$ f\/inite and consider the case
of arbitrary inhomogeneity parameters in this direction. The
fermionic basis will not be af\/fected~\cite{HGSIII}. There are only
two transcendental functions  $\rho$ and $\omega$ which ``absorb''
the whole physical information and appear in the determinants in
analogy with the situation with free fermions.
 In~\cite{HGSIII} these functions were
represented in terms of the dif\/ferential of the second kind in the
theory of deformed Abelian integrals for f\/inite $\mathbf{n}$. These
functions can be also obtained within the TBA approach, \cite{BG}
which makes it possible to take the Trotter limit
$\mathbf{n}\rightarrow\infty$. In the recent papers
\cite{JMSsG1,JMSsG2} the fermionic basis was used in order to study
the one-point functions of the sine-Gordon model on a cylinder. In
this case the transcendental functions must be modif\/ied.

Coming back to the similarity with the CFT, let us emphasize that
the construction of the fermionic basis bears some similarity with
the construction of the descendants in CFT via the action of the
Virasoro generators on the primary f\/ields. In the paper
\cite{HGSIV} we tried to make this similarity more explicit. Namely,
we related the main building blocks of CFT -- the three-point
correlators to the scaling limit of a special partition functions of
the six-vertex model constructed with the help of the fermionic
operators. More concretely, we believe that the following conjecture
is true. There exist scaling limits $\taub^*$, $\betab^*$,
$\gammab^*$ of the operators $\tb ^*$, $\bb ^*$, $\cb ^*$ and
scaling limits $\rho^{\text{sc}}$ and $\omega^{\text{sc}}$ of the
functions $\rho$ and $\omega$ as well. The operator $\taub^*$
generates Zamolodchikov's integrals of motion mentioned above, while
the asymptotic expansion of the function $\rho^{\text{sc}}$ with respect
to its spectral parameter generates their vacuum expectation values.
Another conjecture was that the asymptotic expansion of
$\omega^{\text{sc}}$ with respect to the spectral parameters
describes the expectation values of descendants for CFT with central
charge \mbox{$c=1-6\nu^2/(1-\nu)$} where $\nu$ is related to the
deformation parameter $q=e^{\pi i \nu}$. Equivalently, one takes the
six-vertex model on a cylinder for some special boundary conditions
and considers the scaling limit of the corresponding normalized
partition function with inserted quasi-local operator $q^{2\al
S(0)}\mathcal{O}$. On the other hand one computes a normalized
three-point function in the CFT on an inf\/inite cylinder with
 Virasoro descendants of some primary f\/ield
$\phi_{\al}$ of conformal dimension $\Delta_{\al}$ inserted at the
origin and two primary f\/ields $\phi_{\pm}$ inserted at $+\infty$
and $-\infty$ respectively. The conjecture states that for any
quasi-local operator one can f\/ind a corresponding descendant in
such a way that the above partition function and the CFT three-point
function are equal. In order to state this equality we used the
fermionic basis for the quasi-local operators mentioned above.

In \cite{HGSIV} we established such a correspondence between the
fermionic basis and the Virasoro generators only up to the level 6.
It corresponds to bilinear combinations of the ``Fourier-modes''
$\betab^*_j$ and $\gammab^*_j$ with odd index $j$. For level 8 we
have 5 Virasoro descendants $\mathbf{l}_{-2}^4$,
$\mathbf{l}_{-4}\mathbf{l}_{-2}^2$, $\mathbf{l}_{-4}^2$,
$\mathbf{l}_{-6}\mathbf{l}_{-2}$, $\mathbf{l}_{-8}$. As for the
fermionic basis, we also f\/ind f\/ive linearly independent
combinations of the creation operators $\betab^*_j$ and
$\gammab^*_j$. One of them is quartic:
$\betab^*_1\betab^*_3\gammab^*_3\gammab^*_1$. Unfortunately, we
could not uniquely f\/ix the transformation matrix between the
fermionic basis and the basis generated by the Virasoro descendants.
We thought that it could be done if we would insert the simplest
excitation corresponding to $L_{-1}\phi_{\pm}$  at $\pm\infty$
instead of the primary f\/ields $\phi_{\pm}$ themselves. The
dif\/ference between the ``global'' Virasoro generators $L_{n}$ and
the ``local'' ones $\mathbf{l}_{n}$ was explained in \cite{HGSIV}.
It is also discussed in the next section.

The original motivation of the present paper was to f\/ill this gap.
It turned out, however, that it was not enough to consider only the
simplest excitation. The above uncertainty still remained in this
case. One has to take at least the next excitation, namely, that one
corresponding to the descendants of the second level
$L_{-1}^2\phi_{\pm}$ and $L_{-2}\phi_{\pm}$ in order to f\/ix the
unknown elements of the transformation matrix. All these elements
are certain rational functions of the central charge $c$ and the
conformal dimension of the primary f\/ield
$\Delta_{\al}$.\footnote{Look at the formulae (12.4) of \cite{HGSIV}
or (\ref{evenodd}) and (\ref{resM817})--(\ref{resM8det}).} They do
not depend on the conformal dimensions $\Delta_{\pm}$ of the primary
f\/ields $\phi_{\pm}$. They are also independent of the choice of
the excitation at $\pm\infty$. This is rather strong condition. We
are still unable to prove it for arbitrary excitation. We think that
it is interesting to consider excitations also independently of the
above concrete problem. Therefore we consider them in a more general
setting and come to the solution of the level 8 problem in the very
end.

The paper is organized as follows. In Section~\ref{section2} we
remind the reader about the main results of the paper \cite{HGSIV}.
Section~\ref{section3} is devoted to the TBA approach for the
excited states. We derive the equation for the auxiliary function
$\Theta$ and def\/ine the function $\omegab$ in
Section~\ref{section4}. In Section~\ref{section5} we discuss the
relation to certain CFT three-point functions. We discuss the
solution of the above level 8 problem in Section~\ref{section6}. In
Appendix~\ref{AppendixA} we show several leading terms of the
asymptotic expansion of the function $\Psi$ def\/ined in
Section~\ref{section3} and discuss its relation to the integrals of
motion. In Appendices~\ref{AppendixB} and~\ref{AppendixC} several
leading terms for the asymptotic expansions of the functions
$\bar{F}$, $x^{\pm}$, $\Theta$ are explicitly shown.

\section{Reminder of basic results of \cite{HGSIV}}\label{section2}

As was mentioned in the Introduction, in the paper \cite{HGSIV} some
specif\/ic connection between the conformal f\/ield theory (CFT) with
the central
charge $c=1-6\nu^2/(1-\nu)$ and the XXZ model with the deformation parameter $q=e^{\pi i \nu}$ was established\footnote{Usually we take $\nu$ in the region $\frac12<\nu<1$ called
in \cite{BLZII} a ``semi-classical domain''. The region $0<\nu\le
\frac12$ demands more accurate treatment.}. More precisely, the
following relation was found to be valid with the left hand side
containing the CFT data and the right hand side containing the
lattice model data
\begin{gather}
 \frac{\langle\Delta_-|P_{\al}\bigl(\{\mathbf{l}_{-k}\}\bigr)
\phi_{\al}(0)|\Delta_+\rangle}{\langle\Delta_-|\phi_{\al}(0)|\Delta_+\rangle}=
\lim_{\mathbf{n}\to\infty,\; a\to 0, \; \mathbf{n}a=2\pi R}
Z^{\kappa, s}\bigl\{q^{2\al S(0)}\mathcal{O}\bigr\}.
\label{cft-6v}
\end{gather}
The left hand side means a normalized three-point function of the
CFT def\/ined on a cylinder of radius $R$ parameterized by a complex
variable $z=x+iy$ with spacial coordinate $x$: $-\infty<x<\infty$
and coordinate in the Matsubara direction $y$: $-\pi R<y<\pi R$. The
equivalence of the points $x\pm \pi i R$ is implied. At the origin
$z=0$ some descendant f\/ield is inserted which is given by some
polynomial $P_{\al}(\{\mathbf{l}_{-k}\})$ of Virasoro generators
$\mathbf{l}_{-k}$, $k>0$ acting on the primary f\/ield
$\phi_{\al}(z)$ with conformal dimension
\begin{gather}
\Delta_{\al}=\frac{\nu^2\al(\al-2)}{4(1-\nu)}.
\label{Deltaal}
\end{gather}
We called these Virasoro generators ``local'' in a sense that they are def\/ined in vicinity of $z=0$ with the corresponding energy-momentum tensor
\begin{gather*}
T(z) = \sum_{n=-\infty}^{\infty}\mathbf{l}_{n} z^{-n-2}.
\end{gather*}
The bra- and ket-states $|\Delta_+\rangle$ and $\langle\Delta_-|$  are to be def\/ined through two primary f\/ields $\phi_{\pm}$ with conformal dimensions $\Delta_{\pm}$ being inserted at $x\rightarrow\pm\infty$ in such a way that
$L_n|\Delta_+ \rangle = \delta_{n,0}\Delta_+|\Delta_+ \rangle$, $n\ge 0$
when $x=\infty$ and $\langle\Delta_-|L_n = \delta_{n,0}\Delta_-\langle\Delta_-|$, $n\le 0$ at $x=-\infty$.
We called the Virasoro gene\-ra\-tors $L_n$ ``global''. They correspond to the expansion obtained via the conformal transformation $z\rightarrow e^{-z/R}$
\begin{gather*}
T(z) = \frac{1}{R^2}\left(\sum_{n=-\infty}^{\infty}L_{n} e^{\frac{nz}R} - \frac{c}{24}\right).
\end{gather*}
In \cite{Zam} Alexander Zamolodchikov introduced the local integrals of motion which act on local operators as
\begin{gather*}
(\mathbf{i}_{2n-1}O)(w)
=\int _{{C_w}}\frac {d z}{2\pi i}h_{2n}(z)O (w)
\qquad (n\ge 1),
%\label{i}
\end{gather*}
where the densities $h_{2n}(z)$ are certain
descendants of the identity operator $I$.
An important property is that
\begin{gather}
\langle\Delta_- |\mathbf{i}_{2n-1}\bigl(O(z)\bigr)
|\Delta_+\rangle
=
(I^+_{2n-1}-I^-_{2n-1})
\langle\Delta_-| O(z) |\Delta_+\rangle,
\label{Int-m}
\end{gather}
where $I^{\pm}_{2n-1}$ denote the vacuum eigenvalues of the local
integrals of motion on the Verma module with conformal dimension
$\Delta_\pm$. The Verma module is spanned by the elements
\begin{gather}
\mathbf{i}_{2k_1-1}\cdots \mathbf{i}_{2k_p-1}
\mathbf{l}_{-2l_1}\cdots \mathbf{l}_{-2l_q}(\phi _{\al}(0)) .
\label{basis}
\end{gather}
In case when $\Delta_+=\Delta_-$ the space is spanned by the even Virasoro generators $\{{\bf l}_{-2n}\}_{n\geq1}$.

In order to describe the right hand side of (\ref{cft-6v}) we need the fermionic basis constructed in~\cite{HGSI,HGSII}
via certain creation operators.
These creation operators called $\tb ^*$, $\bb ^*$, $\cb ^*$ together with the annihilation operators called $\bb$, $\cb$ act in the space\footnote{The problem of constructing the annihilation operator corresponding
 to the creation operator $\tb^*$ was discussed
in the paper \cite{BG} but was not solved completely.}
\begin{gather*}
\mathcal{W}^{(\al)}
=\bigoplus\limits _{s=-\infty}^{\infty}\mathcal{W}_{\al-s,s},
\end{gather*}
where $\mathcal{W}_{\al-s,s}$ is the subspace of quasi-local operators of the spin $s$ with the shifted $\al$-parameter.
They all are def\/ined as formal power series of $\z^2-1$ and have the block structure
\begin{alignat*}{3}
&\tb ^*(\z):\ \ &&  \mathcal{W}_{\al
-s,s}\ \to\ \mathcal{W}_{\al-s,s} ,& \\ %\label{blocks}\\
 & \bb ^*(\z) ,\cb (\z)
 :\  \ && \mathcal{W}_{\al-s+1,s-1}\ \to\ \mathcal{W}_{\al-s,s} ,& \nonumber\\
& \cb ^*(\z), \bb(\z) :\ \ && \mathcal{W}_{\al-s-1,s+1}\ \to\ \mathcal{W}_{\al-s,s} .& \nonumber
\end{alignat*}
The operator $\tb ^*(\z)$ plays the role of a generating function of the commuting integrals of motion. In a sense it is bosonic. It commutes with all fermionic operators $\bb(\z)$, $\cb(\z)$ and $\bb^*(\z)$, $\cb^*(\z)$ which obey canonical anti-commutation relations
\begin{gather}
 \bigl[  \cb(\xi),\cb^*(\z)  \bigr]_+=\psi (\xi/\z,\al),\qquad
\bigl[  \bb(\xi),\bb^*(\z)  \bigr]_+=-\psi (\z/\xi,\al)\label{comrel}
\end{gather}
with
\begin{gather*}
\psi (\z,\al)=\frac1{2} \z^{\al} \frac{\z^2 +1}{\z^2-1}.
\end{gather*}
The annihilation operators
$\bb$ and $\cb$ ``kill'' the lattice ``primary f\/ield'' $q^{2\al S(0)}$
\begin{gather*}
 \bb(\z)\big(q^{2\al S(0)}\big)=0,\qquad  \cb(\z)\big(q^{2\al S(0)}\big)=0,\qquad
S(k)=\frac12\sum_{j=-\infty}^k\sigma^3_j.
\end{gather*}
The space of states is generated via the multiple action of the creation operators $\tb ^*(\z)$, $\bb^*(\z)$, $\cb^*(\z)$ on the
``primary f\/ield'' $q^{2\al S(0)}$. In this way one can obtain the fermionic
basis. The completeness of this basis was proved in the paper~\cite{complet}.

In the right hand side of equation~(\ref{cft-6v}) we take the scaling limit of a normalized partition function of the six-vertex model
on a cylinder with insertion of a quasi-local operator $q^{2\al S(0)}\mathcal{O}$
\begin{gather}
 Z^{\kappa, s}
\bigl\{q^{2\al S(0)}\mathcal{O}
\bigr\}=
\frac{\Tr _{\mathrm{S}}\Tr _{\mathbf{M}}\Bigl(Y_\mathbf{M}^{(-s)}T_{\mathrm{S},\mathbf{M}}\ q^{2\kappa S}
\ \mathbf{b}^*_{\infty,s-1}\cdots \mathbf{b}^*_{\infty, 0}
\bigl(
q^{2 \al S(0)}\mathcal{O}\bigr)\Bigr)}
{\Tr _{\mathrm{S}}\Tr _{\mathbf{M}}\Bigl(Y_\mathbf{M}^{(-s)}T_{\mathrm{S},\mathbf{M}}\ q^{2\kappa S}
\ \mathbf{b}^*_{\infty,s-1}\cdots \mathbf{b}^*_{\infty, 0}
\bigl(
q^{2 \al S(0)}\bigr)\Bigr)},
\label{Zkappa}
\end{gather}
where the operators $\bb ^*_{\infty,j}$  are def\/ined through\footnote{Actually, later we discuss the Fateev--Dotsenko condition~(\ref{kappa-prime}) fulf\/illed
for the case $\kappa'=\kappa$. In this case the parameters $\al$ and $s$ are constrained. We will be
mostly interested in the case when $0<\al<2$. Then we need $s<0$ and
another def\/inition of the functional (\ref{Zkappa}) is necessary. It can be done through the
replacement of operators $\bb ^*_{\infty,j}$ by $\bc ^*_{\infty,j}$. This choice was taken
in \cite{JMSsG2} with identif\/ication of the notation there $\bc^*_{\text{screen},-j} =
\bc ^*_{\infty,j}$.}
a singular part when $\z\rightarrow 0$ and
$\bb^* _{\mathrm{reg}}$ is a regular one
\begin{gather*}
\z^{-\al} \bb ^*(\z)(X)=
\sum\limits _{j=0}^{s-1}\z ^{-2j} \bb ^*_{\infty,j}(X)+
\z^{-\al} \bb^* _{\mathrm{reg}}(\z)(X),\qquad X\in\mathcal{W}_{\al-s+1,  s-1}
%\mathrm{spin}(X)=s-1
\end{gather*}
for some operator $X$ of the spin $s-1$, $s>0$ where $ \z^{-\al} \bb^* _{\mathrm{reg}}(\z)(X)$  vanishes at zero.
The monodromy matrix $T_{\mathrm{S},\mathbf{M}}$ is def\/ined via the universal $R$-matrix of $U_q(\slth)$ on the tensor
product of two evaluation representations where the f\/irst one~$\frak{H}_{\mathrm{S}}$ corresponds to the inf\/inite
lattice direction and the second one $\frak{H}_{\mathbf{M}}$ corresponds to the Matsubara direction
\begin{gather*}
T_{\mathrm{S},\mathbf{M}}=\raisebox{.7cm}{$\curvearrowright $}
\hskip -.75cm\prod\limits_{j=-\infty}^{\infty}
T_{j,\mathbf{M}},\qquad
T_{j,\mathbf{M}}\equiv T_{j,\mathbf{M}}(1),
\qquad T_{j,\mathbf{M}}(\z)=\raisebox{.7cm}{$\curvearrowleft $}
\hskip -.6cm\prod\limits_{\mathbf{m=1}}^{\mathbf{n}}
L_{j,\mathbf{m}}(\z)
\end{gather*}
with the standard $L$-operator of the six vertex model
\begin{gather*}
L_{j,\mathbf{m}}(\zeta)
=q^{-\frac 1 2\sigma ^3_j\sigma ^3_\mathbf{m}}
-\z ^2 q^{\frac 1 2\sigma ^3_j\sigma ^3_\mathbf{m}}
-\z \big(q-q^{-1}\big) (\sigma ^+_j\sigma ^-_\mathbf{m}+\sigma ^-_j\sigma ^+_\mathbf{m}) .
\end{gather*}

The ``screening operator'' $Y_\mathbf{M}^{(-s)}$ carries spin $-s$.
As was discussed in \cite{HGSIV}, the functional (\ref{Zkappa}) does
not depend on the concrete choice of the screening operator under
rather mild conditions. Due to common wisdom, in case of an
inf\/inite lattice, one can change boundary conditions and, instead
of taking the traces in the right hand side of (\ref{Zkappa}),
insert two one-dimensional projectors $|\kappa\rangle\langle\kappa|$
and $|\kappa+\al-s,s\rangle\langle\kappa+\al-s,s|$ at the boundary,
where $|\kappa\rangle$ is eigenvector of the transfer matrix
$T_\mathbf{M}(\z,\kappa)=\text{Tr}_j\bigl(T_{j,\mathbf{M}}q^{\kappa\sigma^3_j}\bigr)$
with maximal eigenvalue $T(\z,\kappa)$ in the zero spin sector and
where the eigenvector $|\kappa+\al-s,s\rangle$ corresponds to the
maximal eigenvalue $T(\z,\kappa+\al-s,s)$ of the transfer matrix
$T_\mathbf{M}(\z,\kappa+\al-s)$ in the sector with spin $s$. The
twist parameter $\kappa$ plays the role of the magnetic f\/ield. So,
we can perform the following substitution
\begin{gather*}
Z^{\kappa, s}
\bigl\{q^{2\al S(0)}\mathcal{O}
\bigr\}\rightarrow
\frac{\langle\kappa+\al-s,s|T_{\mathrm{S},\mathbf{M}}\ q^{2\kappa S}
\ \mathbf{b}^*_{\infty,s-1}\cdots \mathbf{b}^*_{\infty, 0}
\bigl(
q^{2 \al S(0)}\mathcal{O}\bigr)|\kappa\rangle}
{\langle\kappa+\al-s,s|T_{\mathrm{S},\mathbf{M}}\ q^{2\kappa S}
\ \mathbf{b}^*_{\infty,s-1}\cdots \mathbf{b}^*_{\infty, 0}
\bigl(
q^{2 \al S(0)}\bigr)|\kappa\rangle},
%\label{Zkappa1}
\end{gather*}
which does not af\/fect the answer for the case of inf\/inite lattice if
\begin{gather*}
\langle \kappa |Y^{(-s)}_{\mathbf{M}}
|\kappa+\al-s,s\rangle\ne 0.
\end{gather*}

The theorem proved by Jimbo, Miwa and Smirnov \cite{HGSIII} claims\footnote{Actually, in \cite{HGSIII}
the statement was proved for the case $s=0$ but as was discussed in \cite{HGSIV},
this statement can be proved for $s\neq 0$ also.} that
\begin{gather}
Z^{\kappa,s}\bigl\{\tb^*(\z)(X)\bigr\}
=2\rho(\z|\kappa,\kappa+\al,s)Z^{\kappa,s}\{X\} ,\label{JMS}\\
Z^{\kappa,s}\bigl\{\bb^*(\z)(X)\bigr\}
=\frac 1{2\pi i}\oint_{\Gamma}
\omega (\z,\xi|\kappa,\al,s)
Z^{\kappa,s}\bigl\{\cb(\xi)(X)\bigr\}
\frac{d\xi^2}{\xi^2} ,\nonumber\\%\label{mainb}\\
Z^{\kappa,s}\bigl\{\cb^*(\z)(X)\bigr\}
=-\frac 1 {2\pi i}\oint_{\Gamma}
\omega (\xi,\z|\kappa,\al,s)
Z^{\kappa,s}\bigl\{\bb(\xi)(X)\bigr\}
\frac{d\xi^2}{\xi^2} ,
\nonumber%\label{mainc}
\end{gather}
where the contour $\Gamma$  goes around all the singularities of
the integrand except $\xi^2=\z^2$.
The direct consequence of the above theorem and the anti-commutation relations (\ref{comrel}) is
the determinant formula
\begin{gather}
 Z^{\kappa,s}\bigl\{\tb^*(\z^0_1)\cdots \tb^*(\z^0_p)
\bb^*(\z^+_1)\cdots \bb^*(\z^+_r)
\cb^*(\z^-_r)\cdots \cb^*(\z^-_1)\bigl(q^{2\al  S(0)}\bigr)\bigr\}
\label{det}
\\
 \qquad {} =\prod\limits _{i=1}^p 2\rho (\z _i^{0}|\kappa,\kappa+\al,s)\times
\det \big(\omega(\z^+_i,\z ^-_j|\kappa,\al,s)
\big)_{i,j=1,\dots, r}.\nonumber
\end{gather}
The functions $\rho$ and $\om$ are completely def\/ined by the Matsubara data. The function $\rho$ is
the ratio of two eigenvalues of the transfer matrix
\begin{gather}
\rho(\z|\kappa+\al-s,s)=\frac{T(\z,\kappa+\al-s,s)}{T(\z,\kappa)}.
\label{rho}
\end{gather}
We will come to the def\/inition of the function $\omega$ in Section~\ref{section4} in more general case
of presence of the excited states.

The scaling limit in the Matsubara direction means
\begin{gather}
 \mathbf{n}\to\infty, \qquad a\to 0,\qquad  \mathbf{n}a=2\pi R,
\label{scaling}
\end{gather}
where the radius of the cylinder $R$ is f\/ixed. Simultaneously one should rescale the spectral parameter
\begin{gather}
\z=\la {\bar a}^{\nu}, \qquad \bar a = Ca \label{zetalambda}
\end{gather}
with some f\/ine-tuning constant $C$. One of the most important points
of \cite{HGSIV} was to def\/ine the scaling limits of $\rho$ and $\om$
\begin{gather}
 \rhob(\la|\kappa,\kappa')
=\lim_\mathrm{scaling}
\rho(\la \bar{a}^{\nu}|\kappa,\al,s) ,
\label{scalfun}\\
  \omegab(\la,\mu|\kappa,\kappa ',\al)
= \frac{1}{4} \lim_\mathrm{scaling}
\omega(\la \bar{a}^\nu,\mu
\bar{a}^\nu|\kappa,\al,s)
\nonumber
\end{gather}
where $\kappa'$ is def\/ined through an analogue of the Dotsenko--Fateev condition \cite{DF}
\begin{gather}
\kappa'=\kappa +\al+2 {\frac {1-\nu}\nu}s.
\label{kappa-prime}
\end{gather}

The continuum limit can be taken in both directions of the cylinder.

{\it The first conjecture} proposed in \cite{HGSIV} was that the creation operators are well-def\/ined in
the scaling limit for the space direction when $ja=x$ is
f\/inite
\begin{gather*}
 \taub ^*(\la)=
\lim_{a\to 0}\frac12\tb ^*(\la \bar{a}^\nu),
\qquad\betab ^*(\la)=\lim_{a\to 0}\frac12\bb ^*(\la \bar{a}^\nu),\qquad
\gammab ^*(\la)=\lim_{a\to 0}\frac12\cb ^*(\la \bar{a}^\nu)
%\label{operscal}
\end{gather*}
and for the ``primary f\/ield''
\begin{gather*}
\Phi_{\al}(0)=\lim_{a\to 0} q^{2\al S(0)}.
\end{gather*}
Asymptotic expansions at $\la\rightarrow\infty$ look
\begin{gather}
 \log\(\taub^*(\la)\)\simeq
\sum\limits _{j=1}^{\infty}
\taub^*_{2j-1}\la^{-\frac{2j-1}\nu}
 ,
\label{asymp}\\
 \frac 1{\sqrt{\taub^*(\la)}}\betab^*(\la)
\simeq\sum\limits _{j=1}^{\infty}
\betab^*_{2j-1}
\la^{-\frac{2j-1}\nu}
 ,
\qquad\frac 1{\sqrt{\taub^*(\la)}}\gammab^*(\la)
\simeq
\sum\limits _{j=1}^{\infty}\gammab^*_{2j-1}\la^{-\frac{2j-1}\nu}
 .\nonumber
\end{gather}

{\it The next conjecture} based on the bosonisation arguments was that the scaling limit of the space $\mathcal{W}_{\al-s,s}$
belongs to the direct product of two Verma modules
\begin{gather*}
\mathrm{Scaling}\ \mathrm{ limit}
\(\mathcal{W}_{\al-s,s}\)\subset
\mathcal{V}_{\al+2\frac{1-\nu }{\nu}s}
\otimes \overline{\mathcal{V}}_{-\al}
\end{gather*}
and the operators $\taub^*(\la)$, $\betab^*(\la)$, $\gammab^*(\la)$ act non-trivially only on the f\/irst chirality
component and do not touch the second one
\begin{alignat*}{3}
&\taub^*_{2j-1}  :\ \ & &
\mathcal{V}_{\al+2\frac{1-\nu }{\nu}s}\
\to\
\mathcal{V}_{\al+2\frac{1-\nu}{\nu}s}
 ,& \\
&\betab ^*_{2j-1}  :\ \ &&
\mathcal{V}_{\al+2\frac{1-\nu}{\nu}(s-1)} \
\to\
\mathcal{V}_{\al+2\frac{1-\nu}{\nu}s} , & \\
&\gammab^*_{2j-1} :\  \ &&
\mathcal{V}_{\al+2\frac{1-\nu}{\nu}(s+1)} \
\to\
\mathcal{V}_{\al+2\frac{1-\nu}{\nu}s} .
&
\end{alignat*}

Using the results of the paper \cite{BLZII}, we get the asymptotic expansion for $\la\to+\infty$
\begin{gather}
\log\rhob (\la |\kappa,\kappa')
\simeq\sum\limits _{j=1}^{\infty}\la ^{-\frac {2j-1}\nu}
C_j
\big(I_{2j-1}^+-I_{2j-1}^-\big).
\label{asymrho}
\end{gather}
Here the integrals of motion $I_{2j-1}^{\pm}$ are the same as in
(\ref{Int-m}). They correspond to the ``right'' and the ``left''
vacuum and depend on $\kappa$ and $\kappa'$ respectively:
\begin{gather*}
I_{2j-1}^+=I_{2j-1}(\kappa),\qquad I_{2j-1}^-=I_{2j-1}(\kappa').
\end{gather*}
We can identify
\begin{gather*}
\taub^*_{2j-1}=C_j\mathbf{i}_{2j-1}.
\end{gather*}
For the function $\omegab$ we have
\begin{gather}
\omegab(\la,\mu |\kappa,\kappa ',\al)
\simeq
\sqrt{\rhob(\la|\kappa,\kappa')}
\sqrt{\rhob(\mu|\kappa,\kappa')}
\sum\limits_{i,j=1}^{\infty}
\la ^{-\frac{2i-1}\nu}
\mu ^{-\frac{2j-1}\nu}
\omega_{i, j}(\kappa,\kappa ',\al)
\label{asymom}
\end{gather}
when $\la^2,\mu^2\to+\infty$.

The scaling limit of (\ref{det}) is proportional to
\begin{gather}
 Z_R^{\kappa,\kappa '}\bigl\{\taub^*(\la^0_1)\cdots \taub^*(\la^0_p)
\betab^*(\la^+_1)\cdots \betab^*(\la^+_r)
\gammab^*(\la^-_r)\cdots \gammab^*(\la^-_1)
\bigl(\Phi_\al(0)\bigr)
\bigr\}
\label{ZR}
\\
\qquad{} =\prod\limits _{i=1}^p
\rhob(\la _i^{0}|\kappa,\kappa ')
\times
\det
\big(
\omegab(\la^+_i,\la ^-_j|\kappa,\kappa ',\al)
\big)_{i,j=1,\dots, r} .
\nonumber
\end{gather}
If we substitute the expansion (\ref{asymp}) into the left hand side of (\ref{ZR}) and the expansions (\ref{asymrho}), (\ref{asymom}) into the
 right hand side of (\ref{ZR}) we can compare the coef\/f\/icients standing with powers of the spectral parameters and express the functional
$Z_R^{\kappa,\kappa '}$ of any monomial constructed from the modes
$\taub^*_{2j-1}$, $\betab^*_{2j-1}$, $\gammab^*_{2j-1}$ through the
integrals of motion $I_{2n-1}$, coef\/f\/icients $C_n$ and
$\om_{i,j}$. In~\cite{HGSIV} we argued that the eigenvalue
$T(\z,\kappa+\alpha-s,s)$ in the scaling limit (\ref{scaling})
becomes equal to~$T(\z,\kappa')$. This means that, if we choose
$\alpha$ and the spin $s$ in such a way that $\kappa'=\kappa$, then
using~(\ref{rho}) and~(\ref{scalfun}), we get
$\rhob(\la|\kappa,\kappa')=1$. This is an important technical point.
In this case we were able to apply the Wiener--Hopf technique and
obtain the coef\/f\/icients $\om_{i,j}$ as power series in
$\kappa^{-1}$ where $\kappa\to\infty$.

On the other hand, one can evaluate the left hand side of
(\ref{cft-6v}) using the operator product expansion (OPE). In order
to compare with the result obtained by the above lattice method, one
needs to identify the parameters $\kappa$, $\kappa'$, $\alpha$ with
the CFT data. In fact, we already identif\/ied~$\al$ by taking the
formula (\ref{Deltaal}) for the dimension $\Delta_{\al}$ of the
primary f\/ield $\phi_{\al}$. The next step was to take $\Delta_+ =
\Delta_{\kappa+1}$ and $\Delta_- = \Delta_{-\kappa'+1}$.

{\it The most important conjecture} of \cite{HGSIV} was that for the CFT with
central charge $c=1-6\frac {\nu ^2}{1-\nu}$ it is always possible to f\/ind one-to-one correspondence
between a polynomial $P_{\al}(\{\mathbf{l}_{-k}\})$ in the left hand side of (\ref{cft-6v}) and
certain combinations of $\betab^*_{2j-1}$ and $\gammab^*_{2j-1}$.

It is convenient to introduce
\begin{gather*}
\betab _{2m-1}^{*}=
D_{2m-1}(\al)\betab _{2m-1}^{\mathrm{CFT}*},
\qquad \gammab _{2m-1}^{*}=
D_{2m-1}(2-\al)\gammab_{2m-1}^{\mathrm{CFT}*},
\end{gather*}
where
\begin{gather*}
D_{2m-1}(\alpha)=
\frac 1 {\sqrt{i\nu}}
 \Gamma (\nu)^{-\frac{2m-1}\nu}(1-\nu)^{\frac{2m-1}{2}}
  \frac{1}{(m-1)!}
\frac{\Gamma\left(\frac{\alpha}{2}+\frac 1{2\nu}(2m-1)\right)}
{\Gamma\left(\frac{\alpha}{2}+\frac{(1-\nu)
}{2\nu}(2m-1)
\right)}
\end{gather*}
together with even and odd bilinear combinations
\begin{gather}
 \phib _{2m-1,2n-1}^{\mathrm{even}}
=(m+n-1)
\frac{1}{2}
\left(\betab _{2m-1}^{\mathrm{CFT}*}\gammab _{2n-1}^{\mathrm{CFT}*}
+\betab _{2n-1}^{\mathrm{CFT}*}\gammab _{2m-1}^{\mathrm{CFT}*}\right),
\label{evenodd}\\
 \phib_{2m-1,2n-1}^{\mathrm{odd}}=d_{\al}^{-1}
(m+n-1)
\frac{1}{2}
\left(
\betab _{2n-1}^{\mathrm{CFT}*}
\gammab _{2m-1}^{\mathrm{CFT}*}
-
\betab_{2m-1}^{\mathrm{CFT}*}
\gammab _{2n-1}^{\mathrm{CFT}*}
\right),\nonumber
\end{gather}
where
\begin{gather}
 d_\alpha=\frac{\nu(\nu-2)}{\nu-1}(\al-1)=
 {\frac 1 6}\sqrt{(25-c)(24\Delta _{\al}+1-c)}.
\label{dal}
\end{gather}
The Verma module has a basis consisting of the vectors (\ref{basis}).
Conjecturally the same space is spanned by:
\begin{gather}
\mathbf{i}_{2k_1-1}\cdots \mathbf{i}_{2k_p-1}
 \phib _{2m_1-1,2n_1-1}^{\mathrm{even}}
\cdots
\phib _{2m_r-1,2n_r-1}^{\mathrm{even}}
\phib _{2\bar{m}_1-1,2\bar{n}_1-1}^{\mathrm{odd}}
\phib _{2\bar{m}_{\bar r}-1,2\bar{n}_{\bar r}-1}^{\mathrm{odd}}
\bigl(\phi _{\al}\bigr) .
\label{vec2}
\end{gather}
Since
\begin{gather*}
\left[\ \mathbf{l}_0 , \taub^*_{2j-1}\right]
=(2j-1)\taub^*_{2j-1},
\qquad
\left[\ \mathbf{l}_0 , \betab^*_{2i-1}\gammab^*_{2j-1}\right]
=(2i+2j-2)\betab^*_{2i-1}  \gammab^*_{2j-1} ,
\end{gather*}
the descendants of the form (\ref{basis})
and those created by the fermions of the form (\ref{vec2})
must be f\/inite linear combinations of each other if the corresponding levels coincide
\begin{gather*}
\sum_{j=1}^q 2 l_j = \sum_{k=1}^r 2(m_k+n_k-1) + \sum_{\bar k=1}^{\bar r} 2({\bar m}_{\bar k}+{\bar n}_{\bar k}-1).
\end{gather*}
As was discussed above, one can choose the parameters $\al$ and $s$ in such a way that $\kappa'=\kappa$ and $\Delta_+=\Delta_-$.
With this choice we factor out the integrals of motion.
The quotient space of the Verma module modulo the action of
the integrals of motion is spanned by the vectors of the form
\begin{gather*}
\mathbf{l}_{-2l_1}\cdots \mathbf{l}_{-2l_q}(\phi _{\al}(0)).
\end{gather*}

All coef\/f\/icients of the polynomial $P_{\al}(\{\mathbf{l}_{-k}\})$
are independent of $\kappa$. We were able to identify the above basis vectors up to the level 6.
The system of equations is overdetermined but nevertheless it has a solution:
\begin{gather}
 \phib_{1,1}^{\mathrm{even}}\cong
\mathbf{l}_{-2},\qquad
\phib_{1,3}^{\mathrm{even}}\cong\mathbf{l}_{-2}^2
+\frac{2c-32}{9}  \mathbf{l}_{-4},
\qquad
\phib_{1,3}^{\mathrm{odd}}\cong
\frac{2}{3}  \mathbf{l}_{-4},
\label{bcl}\\
 \phib_{1,5}^{\mathrm{even}}\cong
\mathbf{l}_{-2}^3
+\frac{c+2-20\Delta+2c\Delta}{3(\Delta+2)}
\mathbf{l}_{-4}\mathbf{l}_{-2}
\nonumber\\
\phantom{\phib_{1,5}^{\mathrm{even}}\cong}{} +
\frac{-5600+428c-6c^2
+2352\Delta-300c\Delta+12c^2\Delta+896\Delta^2-32c\Delta^2}
{60(\Delta+2)}
\mathbf{l}_{-6},
\nonumber\\
 \phib_{1,5}^{\mathrm{odd}}\cong
\frac{2\Delta}{\Delta+2} \mathbf{l}_{-4}\mathbf{l}_{-2}
+\frac{56-52\Delta-2c+4c\Delta}{5(\Delta+2)}
\mathbf{l}_{-6},
\nonumber\\
 \phib_{3,3}^{\mathrm{even}}\cong
\mathbf{l}_{-2}^3
+\frac{6+3c-76\Delta+4c\Delta}{6(\Delta+2)}  \mathbf{l}_{-2}\mathbf{l}_{-4}
\nonumber\\
\phantom{\phib_{3,3}^{\mathrm{even}}\cong}{}  +
\frac{-6544+498c-5c^2
+2152\Delta-314c\Delta+10c^2\Delta-448\Delta^2+16c\Delta^2}
{60(\Delta+2)}
\mathbf{l}_{-6},
\nonumber
\end{gather}
where for simplicity we took $\Delta\equiv\Delta_{\al}$. We hope it
will not cause some confusion by mixing this~$\Delta$ with the
anisotropy parameter of the XXZ model. In the above formula
(\ref{bcl}) we imply only a weak equivalence ``$\cong$'' between the
left hand side and the right hand side. The sign $\cong$ means that
the left and right hand sides being substituted into the
corresponding expectation value give the same result by acting on
$\Phi_{\al}(0)$ and $\phi_{\al}(0)$ respectively. In other words
$A\cong B$ if and only if
\begin{gather*}
Z_R^{\kappa,\kappa}\bigl(A(\Phi_{\al}(0))\bigr)=
\frac{\langle\Delta_-|B\bigl(\phi_{\al}(0)\bigr)|\Delta_+\rangle}{\langle\Delta_-|\phi_{\al}(0)|\Delta_+\rangle}
%\label{cong}
\end{gather*}
with the functional $Z_R^{\kappa,\kappa'}$  def\/ined in~(\ref{ZR}).
In the next sections we will generalize the scheme of~\cite{HGSIV}
to the case of excited states.

\section{TBA for excited states}\label{section3}%\label{XXZ}

In fact, the Jimbo, Miwa, Smirnov theorem (\ref{JMS}) is valid not
only for eigenvectors~$|\kappa\rangle$, $|\kappa+\al-s,s\rangle$
associated with maximal eigenvalues in the corresponding sectors,
but also for eigenvectors corresponding to excited states. Here we
consider only special excited states, namely, the so-called
particle-hole excitations \cite{GaTadeVeWo}. Some motivation for
this is as follows. In \cite{BLZII} Bazhanov, Lukyanov and
Zamolodchikov formulated several assumptions about the analytical
properties of the eigenvalues of the $Q$-operator with respect to
the square of the spectral parameter~$\z^2$, in particular, that its
zeroes in the complex $\z^2$-plane are either real or occur in
complex conjugated pairs, that an inf\/inite number of zeroes are
real and positive and that there may be only a f\/inite number of
complex or real negative zeroes. An important assumption is that the
real zeroes accumulate towards~$+\infty$ in the variable $\z^2$.
Further, in~\cite{BLZ03} the authors proposed a conjecture that for
the asymptotic analysis, when the parameter $\kappa$ becomes large,
only real positive zeroes corresponding either to the vacuum or to
the excited states are important. The question why real negative
zeroes or complex zeroes are not important for the asymptotic
analysis seems hard. Usually one uses experience coming from the
numerical study and also the arguments based on the analysis of a
small vicinity of the free-fermion point $\nu=1/2$. Counting
arguments play important role here as well. But we do not know any
rigorous proof of this statement. Any further discussion of this
subtle question would lead us beyond the scope of this paper.

Let us start with the case of a lattice with an even f\/inite number
of sites $\bf n$ in the Matsubara direction. The Bethe ansatz
equations (BAE) are usually deduced from the Baxter's
TQ-relation~\cite{baxter} for the eigenvalues $T(\z,\kappa,s)$,
$Q(\z,\kappa,s)$ of the transfer matrix $T_\mathbf{M}(\z,\kappa)$
and the $Q$-operator $Q_\mathbf{M}(\z,\kappa)$ def\/ined by the
formula (3.3) of \cite{HGSIV}
\begin{gather*}
T(\z,\kappa,s)Q(\z,\kappa,s)=d(\z)Q(q\z,\kappa,s)+a(\z)Q\big(q^{-1}\z,\kappa,s\big), \qquad Q(\z,\kappa,s)=\z^{-\kappa+s}A(\z,\kappa,s)
\end{gather*}
with polynomial dependence of the functions $T$ and $A$ on the spectral parameter $\z$
in every spin sector $s$, $0\le s\le \frac{\bf n}2$ and
\begin{gather*}
a(\z)=\big(1-q\z ^2\big)^\mathbf{n},\qquad d(\z)=\big(1-q^{-1}\z^2\big)^\mathbf{n}.
\end{gather*}
In the TBA approach one introduces auxiliary function
\begin{gather*}
\frak{a}(\z,\kappa,s)=\frac{d(\z)Q(q\z ,\kappa,s )}
{a(\z)Q(q ^{-1}\z ,\kappa,s)}
%\label{aux}
\end{gather*}
which satisf\/ies the BAE
\begin{gather}
\frak{a}(\xi_j,\kappa,s)=-1,\qquad j=1,\dots,\frac{\bf n}2-s
\label{BAE}
\end{gather}
with $\frac{\bf n}2-s$ zeros $\xi_j$ of $Q(\z ,\kappa,s)$ called Bethe roots. The BAE in the logarithmic form look
\begin{gather}
\log \frak{a}(\xi_j,\kappa,s)=\pi i m_j,
\label{BAElog}
\end{gather}
where $m_j$ are pairwise non-coinciding odd integers.
The ground state corresponds to $s=0$ and the choice $m_j=2j-1$. The following picture
describes this situation
\begin{gather*}
{{\cdots \circ \ \circ \ \circ\ \circ \ \bullet\
\bullet \ \bullet \ \bullet \ \bullet \cdots} \atop {\text{\tiny \quad -5\; -3\; -1\quad 1\quad 3\quad 5\quad\quad\quad}}}
\end{gather*}
where the black circles correspond to ``particles'' and have positive
odd coordinates $m_j$ while the white circles correspond to ``holes''
and have negative coordinates. The particle-hole excitations can be
got when some of ``particles'' are moved to positions with negative
coordinates. Let us denote $I^{(+,k)}$ an ordered subset of positive
coordinates $I^{(+,k)}_1<\cdots<I^{(+,k)}_k\le \bf n$ which
correspond to the positions of created holes and $I^{(-,k)}$
correspond to negative coordinates of the moved particles\footnote{The case of $s>0$ can be treated similarly to the case
$s=0$. The only dif\/ference is that for the f\/inite $\mathbf{n}$ case
one has less Bethe roots.} $-I^{(-,k)}_k<\cdots<-I^{(-,k)}_1$:
\begin{gather*}
{{\cdots\circ\circ\circ\cdots\cdot\circ \bullet
\circ \cdots\cdot \circ\bullet  \circ\cdots\cdots \circ \ \circ\ \circ \ \bullet\
\bullet \ \bullet\cdots\bullet \circ
\bullet\cdots\bullet \circ\bullet \cdots}
\atop {\text{\tiny $\quad\quad\quad\quad\quad\quad\;\;
-I^{(-,k)}_k\cdots\cdot -I^{(-,k)}_1\cdots\cdot\cdot -5 -3 -1\;\; 1\quad\, 3\quad\, 5\cdots\quad
\cdot\cdot I^{(+,k)}_1\cdots\cdot I^{(+,k)}_k$}}}
\end{gather*}
We will denote the corresponding vector and co-vector $|\kappa;I^{(+,k)},I^{(-,k)}\rangle$ and $\langle\kappa; I^{(+,k)},I^{(-,k)}|$.
Let $\xi^-_r$, $r=1,\dots, k$ be the Bethe roots corresponding to the particles moved into the positions with negative coordinates $-I^{(-,k)}_r$
and $\xi^+_r$ correspond to the holes with positive positions $I^{(+,k)}_r$.

Following the papers \cite{Destri-deVega,BLZII,KBP,Klumper}, we can
rewrite the BAE (\ref{BAE}) in form of the non-linear integral
equation
\begin{gather}
\log \frak{a}(\z,\kappa,s)=-2\pi i\nu(\kappa-s)
+\log \(
\frac{d(\z)}
{a(\z)}   \)
-\int _{\gamma_{(s,k)}} K(\z/\xi)\log \(1+ \frak{a}(\xi,\kappa,s)\)
\frac {d\xi^2}{\xi ^2},
\label{non-linear1}
\end{gather}
where the contour $\gamma_{(s,k)}$ goes around all the Bethe roots $\xi_j$ including
the moved ones $\xi_{j_r}=\xi^-_r$ in the {\it clockwise} direction and the kernel
\begin{gather*}
K(\z,\al)=\frac 1{2\pi i}\Delta _{\z}\psi (\z,\al) ,\qquad
K(\z)=K(\z,0),\qquad \Delta _\z f(\z)=f(q\z )-f(q^{-1}\z ).%\label{defK}
\end{gather*}
For the case of the ground state $s=0$, $k=0$ we have $\gamma_{(0,0)}=\gamma$ with the contour
$\gamma$ used in Section~3 of~\cite{HGSIV}. Let us for simplicity stay with the case $s=0$.
One can rewrite (\ref{non-linear1}) replacing
the contour $\gamma_{(0,k)}$ by $\gamma$ and taking into account the contribution of the
residues from the moved Bethe roots $\xi^-_r$ and holes $\xi^+_r$. There are also other
equations coming from (\ref{BAElog}). Altogether we have the following set of equations
\begin{gather}
 \log \frak{a}(\z,\kappa)=-2\pi i\nu\kappa
+\log \(
\frac{d(\z)}
{a(\z)}   \)+\sum_{r=1}^k\big(g(\z/\xi^+_r)-g(\z/\xi^-_r)\big) \nonumber\\ %\label{non-linear2}\\
\phantom{\log \frak{a}(\z,\kappa)=}{}
-\int _{\gamma} K(\z/\xi)\log \(1+ \frak{a}(\xi,\kappa)\)
\frac {d\xi^2}{\xi ^2},\nonumber\\
 \log \frak{a}(\xi^{\pm}_r,\kappa) =\mp \pi i I^{(\pm,k)}_r,\qquad r=1,\dots,k,
\label{additional}
\end{gather}
where $g(\z)=\log\frac{1-q^2\z^2}{1-q^{-2}\z^2}$.

Now we can consider the scaling limit
and generalize the analysis of Section 10 of \cite{HGSIV}.
Let us start with a small remark. So far, we considered only particles which ``jump'' to the left.
In principle, for the case of f\/inite ${\bf n}$ we should also consider particles which ``jump'' to the right. But
in the scaling limit when ${\bf n}\to\infty$ it is implied that the right tail
of the Bethe ansatz phases in the ground state is inf\/inite. So, the jumps to the right
are irrelevant in the scaling limit. Therefore we will not discuss them here.
Let us introduce the functions
\begin{gather*}
 \Tb(\la,\kappa)=
\lim_{\mathbf{n}\to\infty,\ a\to 0,\
2\pi  R=\mathbf{n}a} T(\la \bar{a}^\nu,\kappa) ,
%\label{TQscal}
\\
 \Qb(\la ,\kappa)
=\lim_{\mathbf{n}\to\infty,\ a\to 0,\
2\pi   R=
\mathbf{n}a} \bar{a}^{\nu\kappa}Q(\la \bar{a} ^\nu,\kappa)
\nonumber
\end{gather*}
and taking into account that for $1/2<\nu<1$ the ratio
$a(\z)/d(\z)\to 1$ in the scaling limit with~$\z$ related to $\la$
as in~(\ref{zetalambda}), we get the scaling limit of the auxiliary
function $\frak{a}$ as
\begin{gather*}
\frak{a}^\mathrm{sc}(\la,\kappa)=\frac{\Qb(\la q,\kappa )}
{\Qb(\la q ^{-1},\kappa)}.
\end{gather*}
We would like to study the asymptotic behavior of $\frak{a}^\mathrm{sc}$ for $\la^2\to\infty$
and $\kappa\to\infty$ in such a way that the variable
\begin{gather*}
t=c(\nu)^{-1}\frac{\lambda^2}{\kappa^{2\nu}} %\label{tlambda}
\end{gather*}
is kept f\/ixed with
\begin{gather}
c(\nu)= \Gamma(\nu)^{-2}e^\delta
\left(\frac{\nu}{2R}\right)^{2\nu},\qquad
\delta=-\nu\log\nu-(1-\nu)\log(1-\nu). \label{cdelta}
\end{gather}
Then the function
\begin{gather*}
F(t,\kappa):=\log\mathfrak{a}^\mathrm{sc}(\la,\kappa)
\end{gather*}
satisf\/ies the equation
\begin{gather}
 F(t,\kappa)-
\int_{1}^\infty K(t/u)F(u,\kappa)\frac{du}{u}
=-2\pi i \nu\kappa+\sum_{r=1}^k\bigl(g(t/t^+_r)-g(t/t^-_r)\bigr)
\label{eqF}\\
\phantom{F(t,\kappa)}{} -
\int_{1}^{e^{i \epsilon}\infty}
K(t/u)\log\big(1+e^{F(u,\kappa)}\big)\frac{du}{u}
+\int_{1}^{e^{-i \epsilon}\infty}
K(t/u)\log\big(1+e^{-F(u,\kappa)}\big)\frac{du}{u},\nonumber
\end{gather}
where $\epsilon$ is a small positive number and
\begin{gather*}
t^{\pm}_r=c(\nu)^{-1}\frac{{\xi^{\pm}_r}^2}{(\kappa\bar a)^{2\nu}}.
\end{gather*}
With a slight abuse of notation we will write
\begin{gather*}
 K(t)=\frac1{2\pi i}\cdot\frac12\left(\frac{tq^2+1}{tq^2-1}
-\frac{tq^{-2}+1}{tq^{-2}-1}\right),\qquad g(t)=\log \frac{tq^2-1}{tq^{-2}-1}.
\end{gather*}
Introducing the resolvent $R(t,u)$
\begin{gather*}
 R(t,u)-\int_1^\infty\frac{dv}vR(t,v)K(v/u)=K(t/u)
\qquad (t,u>1)
\end{gather*}
and
\begin{gather}
 G(t;u,v)=G(t,u)-G(t,v),\nonumber
\\
 G(t,u)=((I+R)g)(t,u)=g(t/u)+\int_1^\infty R(t,v)g(v/u)\frac{dv}{v},
\label{eqonG}
\end{gather}
one can rewrite the equation (\ref{eqF}) as follows
\begin{gather}
 F(t,\kappa)=\kappa F_0(t)+\sum_{r=1}^k G(t;t^+_r,t^-_r)
\nonumber\\
\phantom{F(t,\kappa)=}{} -\int_{1}^{e^{i\epsilon}\infty}R(t,u)\log\bigl(1+e^{F(u,\kappa)}\bigr)\frac{du}{u}
+\int_{1}^{e^{-i\epsilon}\infty}R(t,u)\log\bigl(1+e^{-F(u,\kappa)}\bigr)\frac{du}{u},
\label{eqonF}
\end{gather}
where $t^{\pm}_r$ depend on $\kappa$ and can be def\/ined from the equations
\begin{gather}
F(t^{\pm}_r,\kappa)=\mp\pi i I^{(\pm,k)}_r, \qquad r=1,\dots,k
\label{eqont}
\end{gather}
and $F_0$ is the same as in \cite{HGSIV}. It satisf\/ies the integral equation
\begin{gather}
((I-K)F_0)(t) = -2\pi i \nu,
\label{eqonF0}
\end{gather}
where as in (\ref{eqonG}), the contraction is def\/ined on the interval $[1,\infty)$
\begin{align*}
Kf(t)=\int_{1}^\infty K(t/u)f(u)\frac{du}{u}.
\end{align*}
It follows from the WKB technique \cite{BLZII} that the asymptotic behavior at $t\to\infty$
\begin{gather*}
F_0(t)={\rm const}\cdot t^{\frac{1}{2\nu}}
+O\big(t^{-\frac{1}{2\nu}}\big).
\end{gather*}
Then the equation (\ref{eqonF0}) can be uniquely solved by the Wiener--Hopf factorization technique
\begin{gather*}
 F_0(t)=
\int_{\R-\frac{i}{2\nu}-i0}
dl\,t^{il}S(l)\frac{-if}{l(l+\frac{i}{2\nu})}
\qquad (t>1),
%\label{F0}
\end{gather*}
where
\begin{gather}
f=\frac{1}{2\sqrt{2(1-\nu)}} \label{f}
\end{gather}
and the function
\begin{gather*}
 S(k)=\frac{\Gamma(1+(1-\nu)ik)\Gamma(1/2+i\nu k)}
{\Gamma(1+ik)\sqrt{2\pi(1-\nu)}}e^{i\delta k}
\end{gather*}
with $\delta$ def\/ined in (\ref{cdelta}) satisf\/ies factorization
condition
\begin{gather*}
 1-\hat{K}(k)=S(k)^{-1}S(-k)^{-1}, \qquad \hat{K}(k) =\frac{\sinh(2\nu-1)\pi k}{\sinh\pi k},
%\label{factorcond}
\end{gather*}
where $\hat K$ stands for the Mellin transform of the kernel $K$
\begin{gather*}
\hat{K}(k)=\int_0^\infty K(t)t^{-ik}\frac{dt}{t},\qquad
K(t)=\frac{1}{2\pi}\int_{-\infty}^\infty \hat{K}(k)t^{ik}dk.
\end{gather*}
Below we also use the asymptotic expansion of $S(k)$ at $k\to\infty$
\begin{gather*}
S(k)\simeq 1+\sum_{j=1}^{\infty}S_j (ik)^{-j}.
%\label{asS}
\end{gather*}
For example, the two leading terms look
\begin{gather}
S_1=\frac{(1 + \nu)(2\nu -1)}{24(1 - \nu)\nu},\qquad
S_2=\frac{(1 + \nu)^2(2\nu-1)^2}{1152(1 - \nu)^2\nu^2}.
\label{S1S2}
\end{gather}
The same method leads to
\begin{gather}
R(t,u)=
 \int_{-\infty}^\infty\frac{dl}{2\pi}
\int_{-\infty}^\infty\frac{dm}{2\pi}\;
t^{il}u^{im}S(l)S(m)\hat{K}(m)
\frac{-i}{l+m-i0}\nonumber\\
\phantom{R(t,u)}{} =K(t/u)+
 \int_{-\infty}^\infty\frac{dl}{2\pi}
\int_{-\infty}^\infty\frac{dm}{2\pi}\;
t^{il}u^{im}S(l)S(m)\hat{K}(l)\hat{K}(m)
\frac{-i}{l+m-i0}.\label{R}
\end{gather}

One can solve the equation (\ref{eqonF}) iteratively using the asymptotic expansion
\begin{gather*}
F(t,\kappa) = \sum_{n=0}^{\infty}{\kappa}^{-n+1}F_n(t).
%\label{Fexpan}
\end{gather*}
An important dif\/ference of this expansion with the formula (10.3) of \cite{HGSIV} is that here also
even degrees of $\kappa$ may appear.
Since
\begin{gather*}
-\frac{1}{2\pi i}  u\frac{\partial}{\partial u}g(t/u) = K(t/u) ,
\end{gather*}
we can get the following equation by dif\/ferentiating (\ref{eqonG})
\begin{gather*}
-\frac{1}{2\pi i}  u\frac{\partial}{\partial u}G(t/u) = ((I+R)K)(t,u) = R(t,u)
\end{gather*}
and then using (\ref{R})
\begin{gather*}
G(t,u) =-2\pi i \int_{-\infty}^\infty\frac{dl}{2\pi}
\int_{-\infty}^\infty\frac{dm}{2\pi}\; t^{il}u^{im}
S(l)\hat{K}(l)S(m)\frac{-i}{l+m-i0}\,\frac{-i}{m+i0},
\end{gather*}
where we used the regularization ${m+i0}$ in such a way that
$G(t,u)\to 0$ for $u\to\infty$ but in fact, this does not matter
because $G(t,u)$ enters into the equation~(\ref{eqonF}) only through
the dif\/ference $G(t,u)-G(t,v)$.

One introduces the function $\Psi(l,\kappa)$ which has an asymptotic expansion
\begin{gather}
 \Psi(l,\kappa)\simeq
\sum_{n=0}^\infty\kappa^{-n+1}\Psi_n(l) ,
\qquad
\Psi_0(l)=-\frac{if}{l(l+\frac{i}{2\nu})}
\label{asyPsi}
\end{gather}
related to the function $F(t,\kappa)$ via
\begin{gather}
 F(t,\kappa)=\kappa F_0(t)+
\int_{-\infty}^\infty
dl\,t^{il}S(l)\hat{K}(l)(\Psi(l,\kappa)-
\kappa\Psi_0(l)).\label{F}
\end{gather}

It is convenient to introduce
\begin{gather*}
p:=f\kappa,
\end{gather*}
where $f$ is def\/ined in (\ref{f}) and consider the asymptotic
expansion with respect to~$p$ instead of~$\kappa$.

Using (\ref{F}), one can rewrite (\ref{eqF}) in the following equivalent form
\begin{gather}
\Psi^{(p)}(l,p)-p\Psi^{(p)}_{0}(l)
=\sum_{r=1}^k\hat G(l;t^+_r,t^-_r)
-\frac{i}{p}\Biggl\{
\int_{0}^{-i\infty+\epsilon}\frac{dx}{2\pi}
 \hat{R}(l,e^{ix/p})\log\bigl(1+e^{F^{(p)}(e^{ix/p},p)}\bigr)
\nonumber\\
\phantom{\Psi^{(p)}(l,p)-p\Psi^{(p)}_{0}(l)=}{}
+\int_{0}^{i\infty+\epsilon}\frac{dx}{2\pi}
 \hat{R}(l,e^{-ix/p})\log\bigl(1+e^{-F^{(p)}(e^{-ix/p},p)}\bigr)
\Biggr\},\label{diffPsi}
\end{gather}
where $\epsilon$ is a small positive number and by def\/inition
\begin{gather*}
F^{(p)}(t,p) := F(t,\kappa), \qquad \Psi^{(p)}(l,p):=\Psi(l,\kappa), \qquad \Psi^{(p)}_{0}(l):=\frac1f \Psi_0(l)=
-\frac{i}{l(l+\frac{i}{2\nu})}
%\label{identify}
\end{gather*}
and then
\begin{gather*}
\hat{R}(l,e^{ix/p})=
\res_{h}\left[\frac{e^{-h x/p}}{l+h}S(h)\right],\qquad
\hat{G}(l;y,z)
=-\res_{h}\left[\frac{e^{-hy/p}-e^{-hz/p}}{h(l+h)}S(h)\right],
%\label{R-Ghat}
\end{gather*}
where $\res_{h}$ is the coef\/f\/icient at $h^{-1}$ in the expansion at $h=\infty$.
Taking
\begin{gather*}
 F^{(p)}(e^{ix/p},p)=-2\pi\left(x-\bar{F}(x,p)\right),
\end{gather*}
we obtain the Taylor series at $x=0$
\begin{gather}
 \bar{F}(x,p)= x+\res_{h}\bigl[e^{-h x/p}S(h)i\Psi^{(p)}(h,p)\bigr].
\label{TayF}
\end{gather}
As the asymptotic series with respect to $p$ at $p\to\infty$, it starts with $p^{-1}$.
The $2k$ parameters $t^{\pm}_r$ participating in (\ref{eqont}) and (\ref{diffPsi}) are functions of $\kappa$ or equivalently of $p$.
We take them in the following form
\begin{gather}
 t^{\pm}_r(\kappa)=e^{\frac{i x^{\pm}_r(p)}{p}},
\qquad
x^{\pm}_r(p)=\sum_{j=0}^{\infty}x^{\pm}_{r,j}p^{-j}.
\label{expanx}
\end{gather}
Then we get from (\ref{diffPsi})
\begin{gather}
\Psi^{(p)}(l,p)=\sum_{r=1}^k\hat G(l;x^+_r(p),x^-_r(p))-\frac{i p}{l(l+\frac{i}{2\nu})}+H(l,p),
\label{Psip}
\end{gather}
where
\begin{gather}
H(l,p)=-\frac{2i}{p}\int_0^\infty\frac{dx}{2\pi}
\left\{\res_{h}\left(\frac{e^{-hx/p}}{l+h}S(h)\right)\right.\nonumber\\
\left.\phantom{H(l,p)=}{}\times
\sum_{n=0}^\infty\frac{1}{n!}\bar{F}(x,p)^n\left(-\frac{\partial}{\partial x}\right)^n\right\}_{\rm even}
\log(1+e^{-2\pi x}).
\label{H}
\end{gather}
Here $\{f(x,\frac{\partial}{\partial x})\}_{\rm
even}=\frac12(f(x,\frac{\partial}{\partial
x})+f(-x,-\frac{\partial}{\partial x}))$. The parameters
$x^{\pm}_r(p)$ are to be determined from the condition
\begin{gather}
\bar{F}(x^{\pm}_r(p),p)=x^{\pm}_r(p)\mp \frac{i}2 I^{(\pm,k)}_r
\label{barF}
\end{gather}
which is $O(p^{-1})$ if
\begin{gather}
x^{\pm}_{r,0}=\pm \frac{i}2 I^{(\pm,k)}_r.
\label{x0}
\end{gather}
So, we come to the iterative scheme which allows us to compute $F^{(p)}(t,p)$ to any order of $p^{-1}$.
One can see that the following expansions are consistent with the above equations:
\begin{gather}
 \bar{F}(x,p)=\sum_{j=0}^{\infty}\bar{F}_j(x)p^{-j-1},\label{expanF}\\
 H(l,p)=\sum_{j=0}^{\infty}H_j(l)p^{-j-1},\label{expanH}\\
 \hat{G}(l;y,z)=\sum_{j=0}^{\infty}{\hat{G}}_j(l;y,z)p^{-j-1}\label{expanG}
\end{gather}
and in agreement with (\ref{asyPsi})
\begin{gather}
 \Psi^{(p)}(l,p)=\sum_{j=0}^{\infty}\Psi^{(p)}_j(l)p^{-j+1}.\label{expanPsi}
\end{gather}
Let us explain the very f\/irst iteration step. We f\/irst compute $H_0$ in the expansion (\ref{expanH}) \mbox{using}~(\ref{H}), the formula
\begin{gather*}
 \int_0^\infty\frac{dx}{2\pi}x^m
\left({-}\frac{\partial}{\partial x}\right)^n
\log(1+e^{-2\pi x})
=m!(1-2^{-m-1+n})
\frac{\zeta(m-n+2)}{(2\pi)^{m-n+2}}%\label{cnm}
\end{gather*}
and the fact that only the term with $n=0$ in the sum at the right hand side of (\ref{H})
contributes because $\bar{F}$ is of order $p^{-1}$. So, we easily come to the result
\begin{gather*}
H_0(l) = -\frac{i}{24}.
\end{gather*}
Then we substitute it into (\ref{Psip}) and calculate the leading order of the function $\hat{G}$ taking into account
the condition (\ref{x0}). As a result we obtain a few leading orders of $\Psi^{(p)}$:
\begin{gather*}
\Psi^{(p)}(l,p) = -\frac{i p}{l(l+\frac{i}{2\nu})}+\left(\sum_{r=1}^k\frac{i}{2}(I^{(+,k)}_r+I^{(-,k)}_r)-\frac{i}{24}\right)p^{-1}
+O\big(p^{-2}\big).
%\label{Psi2order}
\end{gather*}
This we substitute into the formula (\ref{TayF}) and get $\bar{F}_0$ from (\ref{expanF})
\begin{gather*}
\bar{F}_0(x) = -\left(\frac{i}{4\nu}+\frac{i  S_1}{2}\right)x^2+S_1\left(\sum_{r=1}^k\frac{i}{2}\big(I^{(+,k)}_r+I^{(-,k)}_r\big)-\frac{i}{24}\right),
\end{gather*}
where $S_1$ is given by (\ref{S1S2}). Then we take the equation (\ref{barF}) up to the order $p^{-1}$
and easily solve it
\begin{gather*}
x^{\pm}_{s,1} = \left(\frac{i}{4\nu}+\frac{i  S_1}{2}\right)\left(\frac{I^{(\pm,k)}_s}{2}\right)^2+
S_1\left(\sum_{r=1}^k\frac{i}{2}(I^{(+,k)}_r+I^{(-,k)}_r)-\frac{i}{24}\right).
\end{gather*}
One can do the second iteration by repeating this procedure. In Appendices~\ref{AppendixA} and~\ref{AppendixB} we show the result of such a calculation for few next orders with respect to~$p^{-1}$.

\section[The function $\omegab$]{The function $\boldsymbol{\omegab}$}\label{section4}

Also we need to generalize the expressions (11.5), (11.6) of \cite{HGSIV} for the function $\omegab$ to the case
of excited states. Still we take the condition $\kappa=\kappa'$ for which we can choose the excited state for the
spin~0 sector and for the sector with spin~$s$ in such a way that $\rho(\la|\kappa,\kappa')=1$. The reasoning here is quite
similar to that one described in Section~4 of~\cite{HGSIV}. Actually, one can start with the similar expression to~(11.1) of~\cite{HGSIV} but with a generalized dressed resolvent $\Rdr$
\begin{gather}
 \omegab(\la,\mu|\kappa,\kappa,\alpha)
=\bigl(
f_\mathrm{left}\;{\star}_k \; f_\mathrm{right}+
f_\mathrm{left}\;{\star}_k\;\Rdr\;
{\star}_k\; f_\mathrm{right}\bigr)(\la,\mu)+\omega_0(\la,\mu|\al),
\label{omega1}
\end{gather}
where\footnote{In comparison with (11.1) of~\cite{HGSIV} we take instead of the function $\psi$ def\/ined in (\ref{comrel}) the
function $\psi_0$. The result does not change if we also change the kernel $K_{\al}\to K_{\al,0}$ as it is done in~(\ref{dress}).}
\begin{gather*}
 f_\mathrm{left}(\la,\mu,\alpha)=
\frac{1}{2\pi i}
\delta_\lambda^-\psi_0(\lambda/\mu,\alpha),
\qquad
f_\mathrm{right}(\la,\mu,\alpha)=\delta_\mu^-\psi_0(\lambda/\mu,\alpha) ,\\
 \delta_{\la}^-f(\la)=f(q\la)-f(\la),\qquad \omega _0(\la,\mu|\al)
=\delta_\lambda^-\delta_\mu^-\Delta ^{-1}_\la
\psi_0(\lambda/\mu,\alpha),\\
 \psi_0(\la,\al)=\frac{\la^{\al}}{\la^2-1} ,\qquad \Delta_{\la}f(\la)=f(q\la)-f(q^{-1}\la)
\end{gather*}
with
\begin{gather*}
\Delta ^{-1}_\la\psi_0(\la,\al)=-VP\int _0^{\infty}\frac{\psi_0(\mu,\al)}{2\nu\bigl(1+\(\la/\mu\)^{\frac 1 \nu}\bigr)}
 \frac{d\mu^2}{2\pi i\mu^2},
%\label{Dinv}
\end{gather*}
where the principal value is taken for the pole $\mu^2=1$. The contraction ${\star}_k$ means
\begin{gather*}
 f\; \star_k\; g=\int_{\tilde\gamma_{(0,k)}} f(\la)g(\la)dm(\la),
\qquad
dm(\la)=\frac{d\la^2}{\la^2(1+\mathfrak{a}^\mathrm{sc}(\la,\kappa))}
\end{gather*}
with the contour $\tilde\gamma_{(0,k)}$ which corresponds to $\gamma_{(0,k)}$ from the equation~(\ref{non-linear1})
but taken for va\-riab\-le~$\la^2$ instead of $\z^2=\la^2{\bar a}^{2\nu}$. We remind the reader that the contour
$\gamma_{(0,k)}$ was taken around all the Bethe roots in the clockwise direction in case of the excited
state with $k$ particles and $k$ holes for the zero spin sector.

The resolvent $\Rdr$ fulf\/ills the integral equation
\begin{gather}
\Rdr-\Rdr  \;  \star_k \; K_{\alpha,0}=K_{\alpha,0}, \qquad K_{\alpha,0}(\la)=\Delta_{\la}\psi_0(\la,\al).
\label{dress}
\end{gather}

Applying a similar trick which we used to derive (\ref{additional}), namely, deforming the contour
$\tilde\gamma_{(0,k)}\to\tilde\gamma_{(0,0)}$ and taking into account additional terms
coming from the residues corresponding to particles and holes, we can obtain
\begin{gather}
 \Rdr(t,u)-R(t,u,\al)
=2\pi i\sum_{r=1}^k\left(\frac{R(t,t^+_r,\al)\Rdr(t^+_r,u)}{F'(t^+_r,\kappa)}-
\frac{R(t,t^-_r,\al)\Rdr(t^-_r,u)}{F'(t^-_r,\kappa)}\right)\nonumber
\\
 \qquad{}
 -\int_1^{e^{i\epsilon}\infty}\frac{R(t,v;\al)\Rdr(v,u)}{1+e^{-F(v,\kappa)}}\frac{dv}{v}
-\int_1^{e^{-i\epsilon}\infty}\frac{R(t,v;\al)\Rdr(v,u)}{1+e^{F(v,\kappa)}}\frac{dv}{v},\label{Rdress}\\
F'(t,\kappa):=t \frac{\partial}{\partial t}F(t,\kappa),
\nonumber
\end{gather}
where as in \cite{HGSIV}, we introduced the ``undressed'' resolvent $R(t,u,\alpha)$ which satisf\/ies the equation
\begin{gather}
R(t,u,\alpha)-\int_1^\infty \frac{dv}{v}
R(v,u,\alpha)K_0(t/v,\alpha)=K_0(t/u,\alpha)\label{Rundress}
\end{gather}
with the kernel
\begin{gather*}
K_0(t,\al)=
\frac1{2\pi i}
\left(\frac{(tq^2)^{\al/2}}{tq^2-1}-\frac{(tq^{-2})^{\al/2}}{tq^{-2}-1}\right)
\end{gather*}
corresponding to the above kernel $K_{\al,0}$. The solution of (\ref{Rundress}) again
can be got by the Wiener--Hopf factorization technique
\begin{gather*}
 R(t,u,\alpha)=K_0(t/u,\alpha)\\
\hphantom{R(t,u,\alpha)=}{}
+\int_{-\infty}^\infty\frac{dl}{2\pi}
\int_{-\infty}^\infty\frac{dm}{2\pi}
t^{il}u^{im}S(l,\alpha)
S(m,2-\alpha)\hat{K}(l,\alpha)\hat{K}(m,2-\alpha)
\frac{-i}{l+m-i0}
\end{gather*}
with the Mellin-transform
\begin{gather*}
 \hat{K}(k,\alpha)=
\frac{\sinh\pi\bigl((2\nu-1)k-\frac{i\alpha}{2}\bigr)}
{\sinh\pi\bigl(k+\frac{i\alpha}{2}\bigr)}
\end{gather*}
corresponding to the kernel $K_0(t,\al)$ and
\begin{gather*}
 S(k,\alpha)=
\frac{\Gamma\bigl(1+(1-\nu)ik-\frac{\alpha}{2}\bigr)
\Gamma\bigl(\frac{1}{2}+i\nu k\bigr)}
{\Gamma\bigl(1+ik-\frac{\alpha}{2}\bigr)
\sqrt{2\pi}(1-\nu)^{(1-\alpha)/2}}e^{i\delta k},\\
1-\hat{K}(k,\alpha)=S(k,\alpha)^{-1}S(-k,2-\alpha)^{-1}.
\end{gather*}
We assume that
\begin{align*}
0<\alpha<2 .
\end{align*}
Now we take the ansatz for $\Rdr(t,u)$
\begin{gather}
 \Rdr(t,u)=
K_0(t/u,\alpha)\label{ansatzRdr}
\\
 \phantom{\Rdr(t,u)=}{}
+\int_{-\infty}^\infty\frac{dl}{2\pi}
\int_{-\infty}^\infty\frac{dm}{2\pi}
t^{il}u^{im}S(l,\alpha)S(m,2-\alpha)\hat{K}(l,\alpha)
\hat{K}(m,2-\alpha)\Theta(l,m|p,\alpha)
\nonumber
\end{gather}
with the asymptotic expansion\footnote{Here $\Theta_n(l,m|\alpha)$ are dif\/ferent from
those introduced in \cite{HGSIV} since we should take into account the contribution of
terms with odd degrees $n$ also.}
 at $p\to\infty$
\begin{gather}
 \Theta(l,m|p,\al)
\simeq
\sum_{n=0}^\infty \Theta_n(l,m|\al)p^{-n} ,
\qquad \Theta_0(l,m|\al)=-\frac{i}{l+m}
\label{asyTheta}
\end{gather}
and substitute it into the equation~(\ref{Rdress}). As a result we get the equation
which allows us to calculate all $\Theta_n$ by iterations
\begin{gather}
 \Theta(l,m|p,\al)-\Theta_0(l,m|\al)=
\frac{1}p\, \res_{l'}\res_{m'}\Biggl[S(l',2-\al)S(m',\al)\Theta(m',m|p,\al)/(l+l')
\nonumber\\
 {}\times\Biggl(-i\sum_{r=1}^k e^{-(l'+m')x^+_r(p)/p}/(\bar{F}'(x^+_r(p),p)-1)
+i\sum_{r=1}^k e^{-(l'+m')x^-_r(p)/p}/(\bar{F}'(x^-_r(p),p)-1)\nonumber\\
{}+2\sum_{n=0}^{\infty}\frac{1}{n!}\int_0^{\infty}dx\left\{e^{-(l'+m')x/p}{\bar{F}(x,p)}^n
\left(-\frac{\partial}{\partial x}\right)^n\right\}_{\rm odd}\frac{1}{1+e^{2\pi x}}\Biggr)\Biggr],
\label{eqonTheta}
\end{gather}
where the odd part $\{f(x,\frac{\partial}{\partial x})\}_{\rm
odd}=\frac12(f(x,\frac{\partial}{\partial
x})-f(-x,-\frac{\partial}{\partial x}))$
and
\begin{gather*}
\bar{F}'(x,p):=\frac{\partial}{\partial x}\bar{F}(x,p).
\end{gather*}
Performing iterations implies that enough many orders in the expansion of $\bar{F}(x,p)$ and $x^{\pm}(p)$
with respect to $p^{-1}$ were obtained by means of the iteration scheme described in the previous section.
For a few leading terms in the expansion for $\Theta(l,m|p,\al)$ we get
\begin{gather*}
\Theta(l,m|p,\al)=-\frac{i}{l+m} + \left(\frac{1}{24\nu}-\frac{1}{2\nu}\sum_{r=1}^k\big(I^{(+,k)}_r+I^{(-,k)}_r\big)\right)\\
\phantom{\Theta(l,m|p,\al)=}{}\times
\left(-i\nu(l+m)-\frac{1}{2}+\Delta_\alpha\right)p^{-2}+O\big(p^{-3}\big),
%\label{expTheta}
\end{gather*}
where $\Delta_{\al}$ is given by (\ref{Deltaal}). Some other terms of this expansion will be shown in Appendix~\ref{AppendixC} for the case $k=1$.

One can derive relation like (11.5) of \cite{HGSIV} using the form (\ref{omega1}), integral equations (\ref{dress}), (\ref{Rdress})
for the dressed resolvent $\Rdr$ and the ansatz (\ref{ansatzRdr})
\begin{gather}
\omegab(\la,\mu|\kappa,\kappa,\alpha)
\label{omth}
\\
\simeq \frac{1}{2\pi i}
\int_{-\infty}^\infty dl
\int_{-\infty}^\infty dm\;
\tilde{S}(l,\alpha)
\tilde{S}(m,2-\alpha)
\Theta(l+i0,m|p,\alpha)
\left(\frac{e^{\delta+\pi i\nu}\la^2}
{\kappa^{2\nu}c(\nu)}\right)^{il}
\left(\frac{e^{\delta+\pi i\nu}\mu^2}
{\kappa^{2\nu}c(\nu)}\right)^{im},
\nn
\end{gather}
where we returned to the variables $\la$, $\mu$ and
\begin{gather*}
 \tilde{S}(k,\alpha)=
\frac{\Gamma\bigl(-ik+\frac{\alpha}{2}\bigr)
\Gamma\bigl(\frac{1}{2}+i\nu k\bigr)}
{\Gamma\bigl(-i(1-\nu)k+\frac{\alpha}{2}\bigr)
\sqrt{2\pi}(1-\nu)^{(1-\alpha)/2}}.
\end{gather*}
The asymptotic expansion at $\la,\mu\to\infty$ can be obtained by computing the residues
of the functions $\tilde{S}(l,\alpha)$ and $\tilde{S}(m,2-\alpha)$
\begin{gather}
\omegab(\la,\mu|\kappa,\kappa,\alpha)
 \simeq
\sum_{r,s=1}^\infty\!
\frac{1}{r+s-1}
D_{2r-1}(\alpha)D_{2s-1}(2-\alpha)
\la^{-\frac{2r-1}\nu}
\mu ^{-\frac{2s-1}\nu}
\Omega_{2r-1,2s-1}(p,\alpha),\!\!\!
\label{asym-omega}
\end{gather}
where
\begin{gather}
 D_{2n-1}(\alpha)=
\frac 1{\sqrt{i\nu}}
  \Gamma (\nu)^{-\frac{2n-1}\nu}(1-\nu)^{\frac{2n-1}{2}}
%\cdot
\frac{1}{(n-1)!}
\frac{\Gamma\left(\frac{\alpha}{2}+\frac 1{2\nu}(2n-1)\right)}
{\Gamma\left(\frac{\alpha}{2}+\frac{(1-\nu)
}{2\nu}(2n-1)
\right)}
\label{Phi}
\end{gather}
and
\begin{gather}
\Omega_{2r-1,2s-1}(p,\alpha)
 =-\Theta\left(\frac{i(2r-1)}{2\nu} ,\frac{i(2s-1)}{2\nu}
\Bigl|
p,\alpha\right)
%\cdot
\biggl(\frac{r+s-1}{\nu}\biggr)%\cdot
\left(\frac{\sqrt{2}   p \nu}{R}\right)^{2r+2s-2} .
\label{Omega}
\end{gather}
The relations (\ref{omth})--(\ref{Omega}) look the same as (11.5)--(11.7) of \cite{HGSIV}. However there is an important
dif\/ference. It was pointed out in \cite{HGSIV} that the expansion coef\/f\/icients
$\Theta_n\left(\frac{i(2r-1)}{2\nu} ,\frac{i(2s-1)}{2\nu}\Bigl|p,\alpha\right)$ satisfy the so-called
vanishing property i.e.\ for given $r$ and $s$ they vanish starting from $2n=r+s$. It is
equivalent to the fact that the function $\Theta$ is proportional to a polynomial with respect to~$p$.
This is true only for the case of ground state $k=0$ and also for the case $k=1$ with $I^{(+,1)}=I^{(-,1)}=1$.
For both cases all the coef\/f\/icients $\Theta_n$ with odd $n$ vanish. We will see that in both cases the space of the CFT-descendants is one-dimensional.

\section{Relation to the CFT}\label{section5}

Here we would like to study generalization of the relation (\ref{cft-6v}) between the lattice six vertex model and the CFT
to the case of excited states.
As discussed in Section~\ref{section2}, for the case of the ground state we inserted at $\pm\infty$-points of the cylinder the two primary f\/ields $\phi_{\pm}$
with dimensions~$\Delta_{\pm}$. We also identif\/ied $\Delta_+=\Delta_{\kappa+1}$, $\Delta_-=\Delta_{-\kappa'+1}$
and $\Delta_+=\Delta_-$ for the case $\kappa=\kappa'$. The corresponding states were denoted $|\Delta_+\rangle$
and $\langle\Delta_-|$ respectively.
For the six vertex model we introduced in Section~\ref{section3} the states $|\kappa;I^{(+,k)},I^{(-,k)}\rangle$
marked by the two ordered sets~$I^{(+,k)}$ and~$I^{(-,k)}$ of $k$ odd, positive, non-coinciding numbers.
In case of the spin $s$ sector we denote such a state $|\kappa,s;I^{(+,k)},I^{(-,k)}\rangle$.
In the scaling limit we can identify these states with descendants of the primary f\/ields $\phi_{\pm}$ at level
$N=\frac{1}{2}\sum\limits_{r=1}^k(I^{(+,k)}_r+I^{(-,k)}_r)$
\begin{gather}
 |\kappa ;I^{(+,k)},I^{(-,k)}\rangle\ {{\to}_{\text{scal}}}\ {|\Delta_+ ;I^{(+,k)},I^{(-,k)}\rangle}=
\sum\limits_{{n_1\ge\cdots\ge n_m\ge 1}\atop{n_1+\cdots+n_m=N}}\!
A_{n_1,\dots,n_m}^{(I^{(+,k)},I^{(-,k)})} L_{-n_1}\cdots L_{-n_m}\;|\Delta_+\rangle,
\nonumber\\
 \langle \kappa+\al-s,s; I^{(+,k)},I^{(-,k)}|\ {\to}_{\text{scal}}\
{\langle\Delta_- ;I^{(+,k)},I^{(-,k)}|} \nonumber\\
\qquad{} =
\sum\limits_{{n_1\ge\cdots\ge n_m\ge 1}\atop{n_1+\cdots+n_m=N}}
{\bar A}_{n_1,\dots,n_m}^{(I^{(+,k)},I^{(-,k)})} \langle\Delta_-| L_{n_m}\cdots L_{n_1}.\label{states}
\end{gather}
We normalize
\begin{gather}
\langle\Delta_-; I^{(+,k)},I^{(-,k)}|\phi_{\al}(0)|\Delta_+;I^{(+,k)},I^{(-,k)}\rangle =1
\label{norm}
\end{gather}
but we do not demand orthogonality of the states with dif\/ferent sets. In other words, the ``scalar product''
$\langle\Delta_-; I^{(+,k)},I^{(-,k)}|\phi_{\al}(0)|\Delta_+;{I'}^{(+,k)},{I'}^{(-,k)}\rangle$
is not necessarily zero.
So, instead of (\ref{cft-6v}) we take
\begin{gather}
 \sum\limits_{{n_1\ge\cdots\ge n_m\ge 1}\atop{n_1+\cdots+n_m=N}}
\sum\limits_{{n'_1\ge\cdots\ge n'_{m'}\ge 1}\atop{n'_1+\cdots+n'_{m'}=N}}
A_{n_1,\dots,n_m}^{(I^{(+,k)},I^{(-,k)})}
{\bar A}_{n'_1,\dots,n'_{m'}}^{(I^{(+,k)},I^{(-,k)})}\nonumber\\
\qquad{}\times
\langle\Delta_-|L_{n'_{m'}}\cdots L_{n'_1} P_{\al}\bigl(\{\mathbf{l}_{-j}\}\bigr)
\phi_{\al}(0)L_{-n_1}\cdots L_{-n_m}|\Delta_+\rangle\label{cft-6vexcited}\\
 =\lim_{\substack{\mathbf{n}\to\infty,\\ a\to 0,\\ \mathbf{n}a=2\pi R}}
\frac{\langle\kappa+\al-s,s;I^{(+,k)},I^{(-,k)}|T_{\mathrm{S},\mathbf{M}}  q^{2\kappa S}
  \mathbf{b}^*_{\infty,s-1}\cdots \mathbf{b}^*_{\infty, 0}
\bigl(q^{2 \al S(0)}\mathcal{O}\bigr)|\kappa;I^{(+,k)},I^{(-,k)}\rangle}
{\langle\kappa+\al-s,s;I^{(+,k)},I^{(-,k)}|T_{\mathrm{S},\mathbf{M}}  q^{2\kappa S}
  \mathbf{b}^*_{\infty,s-1}\cdots \mathbf{b}^*_{\infty, 0}
\bigl(q^{2 \al S(0)}\bigr)|\kappa;I^{(+,k)},I^{(-,k)}\rangle}.
\nonumber
\end{gather}
In this relation the polynomial $P_{\al}\bigl(\{\mathbf{l}_{-j}\}\bigr)$ does not depend on $\kappa$ and
the choice of the excitation i.e.\ it is independent of $k$ and of the both sets $I^{(\pm,k)}$.
As was pointed out above, the coef\/f\/icients of this polynomial are
rational functions of the conformal charge $c$ and the conformal dimension~$\Delta_{\al}$ of the primary f\/ield
$\phi_{\al}$ only. If the operator
$\mathcal{O}=\betab^{\mathrm{CFT}*}_{2i_1-1}\cdots
\betab^{\mathrm{CFT}*}_{2i_n-1}\gammab^{\mathrm{CFT}*}_{2j_n-1}\cdots
\gammab^{\mathrm{CFT}*}_{2j_1-1}$ then the level of descendants participating in $P_{\al}\bigl(\{\mathbf{l}_{-j}\}\bigr)$ is
$M=2\sum\limits_{l=1}^n (i_l+j_l-1)$.
On the other hand, the coef\/f\/icients $A_{n_1,\dots,n_m}$, $\bar{A}_{n_1,\dots,n_m} $ are
independent of the choice of the operator $\mathcal{O}$. So, our strategy is to take for a given level~$M$ all linear independent operators $\mathcal{O}$ represented in terms of the fermionic basis
modulo integrals of motion and as many dif\/ferent excitations as necessary in order to f\/ix the corresponding polynomials $P_{\al}\bigl(\{\mathbf{l}_{-j}\}\bigr)$ and the coef\/f\/icients\footnote{In fact, not all coef\/f\/icients $A$, $\bar A$ can be f\/ixed but rather some of their products. We checked for one particular case
that one can f\/ix them completely taking $\bar{A}_{n_1,\dots,n_m}=A_{n_1,\dots,n_m}$ but a'priori it is not quite clear to us
why it should be so.} $A_{n_1,\dots,n_m}$, $\bar{A}_{n_1,\dots,n_m}$.
All other relations for this level partially determine further coef\/f\/icients $A$, $\bar A$  and the rest of the equations fulf\/ills automatically.
In~\cite{HGSIV} we were able to f\/ix the polynomials $P_{\al}\bigl(\{\mathbf{l}_{-j}\}\bigr)$ up to
$M=6$ using only the ground state data when $k=0$. It means that
for $M\le 6$ the relation (\ref{cft-6vexcited}) fulf\/ills automatically for any excitation i.e.\
for any $k$ and any two sets $I^{(\pm,k)}$. We checked this for the case $k=1$ with $I^{(+,1)}_1=I^{(-,1)}_1=1$
and $I^{(+,1)}_1=1$, $I^{(-,1)}_1=3$, $I^{(+,1)}_1=3$, $I^{(-,1)}_1=1$.

Starting with $M=8$ the situation changes. We do not have enough equations in order to f\/ix the polynomials $P_{\al}\bigl(\{\mathbf{l}_{-j}\}\bigr)$
if we restrict ourselves with the case of the ground state. We need additional equations involving excitations. In the next section
we will consider in detail the case $M=8$.

Before we go further let us make one remark.
It is interesting to note that the above fermionic basis operators are marked in exactly the same way as the particle-hole excitations, namely, by two ordered sets of $n$ odd, positive integers $I^{(+,n)}=\{2i_1-1<\cdots < 2i_n-1\}$ and
$I^{(-,n)}=\{2j_1-1<\cdots < 2j_n-1\}$. So, we can denote such an operator
\begin{gather}
{\mathcal{O}}_{I^{(+,n)},I^{(-,n)}}=\betab^{\mathrm{CFT}*}_{2i_1-1}\cdots
\betab^{\mathrm{CFT}*}_{2i_n-1}\gammab^{\mathrm{CFT}*}_{2j_n-1}\cdots
\gammab^{\mathrm{CFT}*}_{2j_1-1}.
\label{basisO}
\end{gather}
In case we take this operator in the right hand side of (\ref{cft-6vexcited})
we will use the following shorthand notation
\begin{gather*}
\text{r.h.s.\ of} \  (\ref{cft-6vexcited})  := \langle I^{(+,k)},I^{(-,k)}|\mathcal{O}_{I^{(+,n)},I^{(-,n)}}|I^{(+,k)},I^{(-,k)}\rangle.
%\label{basisnotation}
\end{gather*}
Now let us describe in general how we compute both sides of the relation (\ref{cft-6vexcited}). We start with the right hand side of~(\ref{cft-6vexcited})
determined by the lattice data. We pointed out the fact that the theorem by Jimbo, Miwa, Smirnov works for the case of excited states
also with the same fermionic operators. Only two transcendental functions~$\rho$ and~$\omega$ are sensible to the changes that
happen in the Matsubara direction. It means that we still can apply the Wick theorem and after taking the scaling limit
come to the same determinant formula~(\ref{ZR}) but with a new functional determined by the
 right hand side of~(\ref{cft-6vexcited})
instead of $Z_{R}^{\kappa,\kappa'}$. With our choice of parameters $\al$, $s$, $\kappa$, $\kappa'$ the function $\rhob$ is still~1 and
the function~$\omegab$ is now determined through the asymptotic expansion~(\ref{asym-omega}). So, we come to
\begin{gather}
 \langle I^{(+,k)},I^{(-,k)}|\mathcal{O}_{I^{(+,n)},I^{(-,n)}}|I^{(+,k)},I^{(-,k)}\rangle
=\det \left(\frac{\Omega_{2i_r-1,2j_{r'}-1}(p,\al)}{i_r+j_{r'}-1}
\right)_{r,r'=1,\dots, n},
\label{basisoperator}
\end{gather}
where the function $\Omega$ is def\/ined by (\ref{Omega}).

Now let us proceed to the left hand side of (\ref{cft-6vexcited}) which involves the CFT data. For simplicity let us put the radius of the cylinder $R=1$.
We need to calculate
\begin{gather*}
\langle\Delta_-|L_{n'_{m'}}\cdots L_{n'_1} P_{\al}\bigl(\{\mathbf{l}_{-j}\}\bigr)
\phi_{\al}(0)L_{-n_1}\cdots L_{-n_m}|\Delta_+\rangle
\end{gather*}
with $n_1\ge \cdots \ge n_m\ge 1$, $n'_1\ge \cdots \ge n'_{m'}\ge 1$ and
\begin{gather*}
N=\sum_{j=1}^m n_j =\sum_{j=1}^{m'} n'_j
\end{gather*}
or picking out some monomial with respect to the local Virasoro generators $\mathbf{l}_{-j}$, we need the following value
\begin{gather}
(n'_1,\dots,n'_{m'};a_1,\dots,a_d;n_1,\dots,n_m)\nonumber\\
\qquad{} :=
\langle\Delta_-|L_{n'_{m'}}\cdots L_{n'_1}\; \mathbf{l}_{-2a_d}\cdots \mathbf{l}_{-2a_1}
\phi_{\al}(0)L_{-n_1}\cdots L_{-n_m}|\Delta_+\rangle
\label{defnan}
\end{gather}
with $d$ positive integers $1\le a_1\le\cdots\le a_d$.

In order to compute it we follow the scheme described in Section~6 of~\cite{HGSIV}. First we def\/ine the function
\begin{gather*}
W(z_1,\dots,z_K;w):=\langle\Delta_-|T(z_K)\cdots T(z_1)\phi_{\al}(w)|\Delta_+\rangle,
\qquad K=d+m+m',
\end{gather*}
where $T(z)$ is the energy-momentum tensor as a function of the point $z$ on the cylinder
with the OPEs
\begin{gather*}
T(z)T(w) =
-\frac{c}{12 }\chi'''(z-w)
-2T(w)\chi'(z-w)
+T'(w)\chi(z-w)+O(1) ,
%\label{OPE1}
\\
T(z)\phi_{\al}(w) =
-\Delta_{\al} \phi_\al(w)\chi'(z-w)
+\phi_\al'(w)\chi(z-w)+O(1)
%\label{OPE2}
\end{gather*}
and
\begin{gather}
\chi(z)=\frac12\coth\left(\frac{z}2\right)
=\sum\limits_{n=0}^\infty\frac {B_{2n}}{(2n)!}z^{2n-1},
\label{localchiexpan}
\end{gather}
where $B_{2n}$ are Bernulli numbers. We also need the expansion:
\begin{gather}
\chi(z)=\pm\frac12\pm\sum\limits_{j=1}^{\infty}e^{\mp j z}, \qquad \Re( z)\to\pm\infty.
\label{chiexpan}
\end{gather}
As was discussed above we can use two dif\/ferent expansions for the energy-momentum tensor as well
\begin{itemize}\itemsep=0pt
\item the {\it ``local'' expansion} in vicinity of $z=0$
\begin{gather*}
T(z) = \sum_{n=-\infty}^{\infty}\mathbf{l}_{n} z^{-n-2},
%\label{localexpan}
\end{gather*}
\item the {\it ``global'' expansion} when $\Re( z)\to\pm\infty$
\begin{gather}
T(z) = \sum_{n=-\infty}^{\infty}L_{n} e^{nz} - \frac{c}{24}.
\label{globalexpan}
\end{gather}
\end{itemize}
The action of the local Virasoro generators $\mathbf{l}_{n}$ on a local f\/ield $O(w)$ is def\/ined
through the contour integral
\begin{gather*}
({\bf l}_nO)(w)=\int_{C_w}\frac{dz}{2\pi i}
(z-w)^{n+1}T(z)O(w),
%\label{actionl}
\end{gather*}
where ${C_w}$ encircles the point $w$ anticlockwise.

The conformal Ward--Takahashi identity allows to determine the function $W(z_1,\dots,z_K)$ recursively:
\begin{gather}
W(z_1,\dots,z_K;w)
=-\frac{c}{12}\sum_{j=2}^K\chi'''(z_1-z_j) W(z_2,\dots\overset{j}{\widehat{\phantom{T}}}\dots, z_K;w)
\label{recurs}\\
\qquad{} +
\Biggl\{\sum_{j=2}^K\left(-2\chi'(z_1-z_j)
+(\chi(z_1-z_j)
-\chi(z_1-w))\frac{\partial}{\partial z_j}\right)
-\Delta_\al\chi'(z_1-w)
\nonumber\\
\qquad{}
+(\Delta_+-\Delta_-)\chi(z_1-w)
+\frac{1}{2}(\Delta_++\Delta_-)-\frac{c}{24}\Biggr\} W(z_2,\dots,z_K;w).
\nonumber
\end{gather}
Of course, the term containing the dif\/ference $\Delta_+-\Delta_-$ drops for the case $\Delta_+=\Delta_-$
which is only interesting for us here.

In order to calculate the above object $(n'_1,\dots,n'_{m'};a_1,\dots,a_d;n_1,\dots,n_m)$
given by (\ref{defnan}), we proceed in several steps:
\begin{itemize}\itemsep=0pt
\item[]\hspace*{-5mm}{\it step 1:}  take the recursion (\ref{recurs}) and expand $\chi(z_1-\cdots)$, $\chi'(z_1-\cdots)$,
$\chi''(z_1-\cdots)$, $\chi'''(z_1-\cdots)$ using the expansion (\ref{chiexpan}) for $\Re(z_1)\to -\infty$ and
then having in mind the expansion (\ref{globalexpan}) take there the coef\/f\/icient at $e^{n'_1 z_1}$,
\item[]\hspace*{-5mm}{\it step 2:} repeat this procedure consequently for the variables $z_2,z_3,\dots,z_{m'}$
taking every time the coef\/f\/icients at $e^{n'_2 z_2},e^{n'_3 z_3},\dots, e^{n'_{m'} z_{m'}}$,
\item[]\hspace*{-5mm}{\it step 3:}
similarly we proceed with the next $m$ variables $z_{m'+1},\dots, z_{m'+m}$ taking the
expansion (\ref{chiexpan}) for $\Re(z_{m'+1})\to \infty,\dots, \Re(z_{m'+m})\to\infty$, further using the
recursion (\ref{recurs}) and picking up coef\/f\/icients at $e^{-n_{1} z_{m'+1}},\dots,
e^{-n_{m} z_{m'+m}}$,
\item[]\hspace*{-5mm}{\it step 4:}
now one can easily compute the limit $w\to 0$ and then apply (\ref{recurs})
with respect to the variable $z_{m'+m+1}$, take the local expansion (\ref{localchiexpan}) of
$\chi(z_{m'+m+1})$, $\chi'(z_{m'+m+1})$, $\chi''(z_{m'+m+1})$, $\dots$ and calculate the contour integral
$\int_{C_0} {dz_{m'+m+1}}{z_{m'+m+1}^{-2a_1+1}}\cdots$ with the expression obtained in this way,
\item[]\hspace*{-5mm}{\it step 5:}
repeat the step~4 with respect to the residual variables $z_{m'+m+2},\dots,z_K$ every time calculating
the contour integrals  $\int_{C_0} {dz_{m'+m+2}}{z_{m'+m+2}^{-2a_2+1}}\cdots$ etc. up to the last
integral $\int_{C_0} {dz_{K}}{z_{K}^{-2a_d+1}}\cdots$.
\end{itemize}

In this way we can obtain, for example, for the case $d=0$, $N=2$
\begin{gather}
 (2;\varnothing;2) = \frac{c}2 +4\;(\Delta^2 - \Delta  + \Delta_+),
\label{N2d0}
\\
 (1,1;\varnothing;2) =(2;\varnothing;1,1) = 2(\Delta^3 -\Delta  + 3\Delta_+),\nonumber\\
 (1,1;\varnothing;1,1) = \Delta(\Delta-1)(\Delta^2-\Delta+2)
+4\Delta_+(2\Delta^2-2\Delta+1)+8\Delta_+^2,\nonumber
\end{gather}
where again we used the shorthand notation $\Delta\equiv\Delta_\al$.

For $d=1$, $a_1=1$, $N=2$ we get
\begin{gather}
 (2;1;2) = -\frac13\Delta(\Delta^2-61\Delta+42)+c\left(-\frac16\Delta^2+\frac{17}8\Delta+1\right)
-\frac{c^2}{48}\nonumber\\
\phantom{(2;1;2) =}{}
+\Delta_+\left(4\Delta^2+\frac{35}3\Delta+\frac{c}3+8\right)+4\Delta_+^2,
\nonumber\\
 (1,1;1;2) =(2;1;1,1) =
-\frac1{12}\Delta(\Delta+1)\big(2\Delta^2 +c\Delta -86\Delta+72-7c\big)\nonumber\\
\phantom{(1,1;1;2) =}{}  +\Delta_+\left((2\Delta+3)
\left(\Delta^2+\frac92\Delta+4\right)-\frac{c}4\right)+6\Delta_+^2,\nonumber\\
(1,1;1;1,1)=-\frac1{24}\Delta(\Delta-1)\big(2\Delta^3+(c-50)\Delta^2
-(c+44)\Delta+2c-96\big)\label{N2d1}\\
\hphantom{(1,1;1;1,1)=}{}
 +\frac{\Delta_+}6\bigl(6\Delta^4+32\Delta^3-2(c-71)\Delta^2
+2(c+29)\Delta+48-c\bigr)\nonumber\\
\hphantom{(1,1;1;1,1)=}{}
+\frac{\Delta_+^2}3\big(24\Delta^2+70\Delta+60-c\big)+
8\Delta_+^3.
\nonumber
\end{gather}
And for $d=1$, $a_1=2$, $N=2$:
\begin{gather}
 (2;2;2)=
\Delta\left(\frac{\Delta^2}{60}+\frac{359}{60}\Delta+\frac{1921c}{480}-3\right)
+\frac{1921\Delta\Delta_+}{60},\nonumber\\
 (1,1;2;2)=
\Delta(\Delta+1)\left(\frac{\Delta^2}{120}+\frac{179}{120}\Delta+c-1\right)
+\Delta\left(12\Delta+\frac{1441}{40}\right)\Delta_+,
\label{N2d1a}\\
 (1,1;2;1,1)=
\Delta^2(\Delta-1)\left(\frac{\Delta(\Delta-1)}{240}+\frac{241}{120}\right)+
\Delta\left(\frac{121}{30}\Delta^2+\frac{599}{30}\Delta+\frac{1921}{60}\right)\Delta_+\nonumber\\
\phantom{(1,1;2;1,1)=}{}
+\frac{481\Delta\Delta_+^2}{30}.
\nonumber
\end{gather}

\section[The level $M=8$]{The level $\boldsymbol{M=8}$}\label{section6}

Here we show how the procedure generally described in the previous section
works for the case of level $M=8$. As was pointed out in Introduction,
there are 5 monomials of the local Virasoro generators
which generate linear independent descendant states modulo integrals of
motion. Let us arrange them as a vector
\begin{gather*}
V_{\mathbf{l}}=
\begin{pmatrix}
\mathbf{l}_{-2}^4\vspace{1mm}\\
\mathbf{l}_{-4}\mathbf{l}_{-2}^2\vspace{1mm}\\
\mathbf{l}_{-4}^2\vspace{1mm}\\
\mathbf{l}_{-6}\mathbf{l}_{-2}\vspace{1mm}\\
\mathbf{l}_{-8}
\end{pmatrix}.
%\label{lbasisM8}
\end{gather*}
 As for the fermionic operators (\ref{basisO}), we also have
5 possibilities for $M=8$ which we also take as a vector
\begin{gather*}
V_{\text{fermi}}=
\begin{pmatrix}
\phib_{1,7}^{\mathrm{even}}\vspace{1mm}\\
\phib_{1,7}^{\mathrm{odd}}\vspace{1mm}\\
\phib_{3,5}^{\mathrm{even}}\vspace{1mm}\\
\phib_{3,5}^{\mathrm{odd}}\vspace{1mm}\\
\betab^{\mathrm{CFT}*}_{1}\,\betab^{\mathrm{CFT}*}_{3}\,
\gammab^{\mathrm{CFT}*}_{3}\,\gammab^{\mathrm{CFT}*}_{1}
\end{pmatrix},
%\label{fbasisM8}
\end{gather*}
where we used the even and odd combinations (\ref{evenodd}).
We would like to determine the 5 by 5 transformation matrix $U$
\begin{gather}
V_{\text{fermi}}\cong U  V_{\mathbf{l}},
\label{trnasformM8}
\end{gather}
where the weak equivalence ``$\cong$'' should be understood in the same way as in the
formula (\ref{bcl}) and all the matrix elements of $U$ depend only on the conformal dimension
$\Delta\equiv\Delta_\al$ and central charge $c$.
In \cite{HGSIV} we were not able to uniquely f\/ix this matrix
from the consideration of the ground state case $N=0$.
If we substitute the both sides of (\ref{trnasformM8}) into~(\ref{cft-6v}) where the matrix elements of $U$ in every row
correspond to the coef\/f\/icients in the polynomial $P_{\al}\bigl(\{\mathbf{l}_{-k}\}\bigr)$
and use (\ref{basisoperator}) for $k=0$ then we get
 \begin{gather}
 \frac{\langle\Delta_-|(U V_{\mathbf{l}})\phi_{\al}(0)|\Delta_+\rangle}{\langle\Delta_-|\phi_{\al}(0)|\Delta_+\rangle}=
V^{(0)},
\label{cft-6vM8}
\end{gather}
where
\begin{gather}
V^{(0)}:=
\begin{pmatrix}
(\Omega^{(0)}_{1,7}+\Omega^{(0)}_{7,1})/2\vspace{1mm}\\
(\Omega^{(0)}_{7,1}-\Omega^{(0)}_{1,7})/(2d_\al)\vspace{1mm}\\
(\Omega^{(0)}_{3,5}+\Omega^{(0)}_{5,3})/2\vspace{1mm}\\
(\Omega^{(0)}_{5,3}-\Omega^{(0)}_{3,5})/(2d_\al)\vspace{1mm}\\
\Omega^{(0)}_{1,1}\Omega^{(0)}_{3,3}/3
-\Omega^{(0)}_{1,3}\Omega^{(0)}_{3,1}/4
\end{pmatrix}
\label{V0}
\end{gather}
and $\Omega^{(0)}_{j,j'}$ denote the functions $\Omega_{j,j'}(p,\al)$
given by (\ref{Omega}) which are calculated for the ground state case
$N=0$, $k=0$ as was explained in Section~11 of \cite{HGSIV}.
From the vanishing property follows that all f\/ive equations in
(\ref{cft-6vM8}) are polynomials with respect to $p^2$ of degree~4. In all 5 cases
one of the equations is fulf\/illed automatically.
It means that every equation leaves an one-parametric freedom. In other words,
in every row of $U$ one matrix element is left undetermined.

Therefore in order to f\/ix the matrix $U$ completely we need to involve excitations.
Here we restrict ourselves with the case $k=1$ and $N\le 2$ only.
Let us remind the reader that the descendant level for the excitations
$N=\frac{1}{2}\sum\limits_{r=1}^k(I^{(+,k)}_r+I^{(-,k)}_r)$.
So, for the case $N=1$ we have only one possibility
$I^{(+,1)}=I^{(-,1)}=\{1\}$. As in the ground state case $N=0$, the descendant
space is one-dimensional because there is only one vector
$L_{-1}|\Delta_+\rangle$ here. When we wrote the paper \cite{HGSIV} we
thought that considering this excitation would help us to f\/ix the above
mentioned freedom. As appeared it is not the case because
it does not produce any additional constraints on the transformation matrix $U$. We leave the checking of this fact as an exercise for the reader.

So, we have to consider the case $N=2$. There are two possibilities
$I^{(+,1)}=\{1\}$, $I^{(-,1)}=\{3\}$ and $I^{(+,1)}=\{3\}$, $I^{(-,1)}=\{1\}$ that correspond
to the two-dimensional descendant space spanned by two vectors
$L_{-2}|\Delta_+\rangle$ and $L_{-1}^2|\Delta_+\rangle$.
Now let us consider the states (\ref{states})
\begin{gather}
|\Delta_+ ;\{1\},\{3\}\rangle=
A  L_{-2}|\Delta_+\rangle+B  L_{-1}^2|\Delta_+\rangle,
\label{13}\\
|\Delta_+ ;\{3\},\{1\}\rangle=
C  L_{-2}|\Delta_+\rangle+D  L_{-1}^2|\Delta_+\rangle,
%\label{31}
\nn\\
\langle\Delta_- ;\{1\},\{3\}|=
\bar{A} \langle\Delta_-|L_2+ \bar{B} \langle\Delta_-|L_1^2,
%\label{bar13}
\nn\\
\langle\Delta_- ;\{3\},\{1\}|=
\bar{C} \langle\Delta_-|L_2+ \bar{D} \langle\Delta_-|L_1^2,
%\label{bar31}
\nn
\end{gather}
where we use simpler notation for the coef\/f\/icients
$A^{(\{1\},\{3\})}_2\equiv A$, $A^{(\{1\},\{3\})}_{1,1}\equiv B$, $A^{(\{3\},\{1\})}_{2}\equiv C$, $A^{(\{3\},\{1\})}_{1,1}\equiv D$
and similar for $\bar{A}$, $\bar{B}$, $\bar{C}$, $\bar{D}$. As appeared we can
take them $\bar{A}=A$, $\bar{B}=B$, $\bar{C}=C$, $\bar{D}=D$.

Now we should satisfy the normalization condition (\ref{norm})
\begin{gather*}
\langle\Delta_- ;\{1\},\{3\}|\phi_\al(0)|\Delta_+ ;\{1\},\{3\}\rangle
= \langle\Delta_- ;\{3\},\{1\}|\phi_\al(0)|\Delta_+ ;\{3\},\{1\}\rangle = 1.
\end{gather*}
Substituting the above formulae (\ref{13}) here, using the notation
(\ref{defnan}) and the fact that $(1,1;\varnothing;2)\! =(2;\varnothing;1,1)$,
we get
\begin{gather}
A^2(2;\varnothing;2)+B^2(1,1;\varnothing;1,1)+2AB(1,1;\varnothing;2)\nonumber\\
\qquad{}=
C^2(2;\varnothing;2)+D^2(1,1;\varnothing;1,1)+2CD(1,1;\varnothing;2)=1,
\label{eqonABCD}
\end{gather}
where $(2;\varnothing;2),(1,1;\varnothing;1,1),(1,1;\varnothing;2)$
are given by~(\ref{N2d0}).
Now we can use the f\/irst row of~(\ref{bcl}), the formulae (\ref{basisoperator}) and (\ref{eqonABCD}) in order to obtain
\begin{align}
\begin{pmatrix}
(2;\varnothing;2) & (1,1;\varnothing;1,1) & (1,1;\varnothing;2)\\
(2;1;2) & (1,1;1;1,1) & (1,1;1;2)\\
(2;2;2) & (1,1;2;1,1) & (1,1;2;2)
\end{pmatrix}
\begin{pmatrix}
A^2\\
B^2\\
2AB
\end{pmatrix}
=
\begin{pmatrix}
1\\
\Omega^{(1,3)}_{1,1}\\
\frac34\bigl(\Omega^{(1,3)}_{3,1}-\Omega^{(1,3)}_{1,3}\bigr)/d_\al
\end{pmatrix}.
\label{eqonAB}
\end{align}
Here we denoted $\Omega^{(1,3)}_{j,j'}$ the function $\Omega_{j,j'}(p,\al)$
given by the formula (\ref{Omega}) where
\begin{gather*}
\Theta(ij/(2\nu),ij'/(2\nu)|p,\al)
\end{gather*}
is determined by the equation (\ref{eqonTheta}) with $k=1$ and
$I^{(+,1)}=\{1\}$, $I^{(-,1)}=\{3\}$ while $d_\al$ is taken from~(\ref{dal}).
The matrix elements of the second and the third row
in the left hand side of~(\ref{eqonAB}) are given by~(\ref{N2d1}) and~(\ref{N2d1a}) respectively.
The matrix equation~(\ref{eqonAB}) can be solved
with respect to $A$ and $B$ by inverting the 3 by 3 matrix. These equations are
overdetermined. The f\/irst two of them give~$A^2$, $B^2$ and we can check that the last equation which gives~$2AB$ is fulf\/illed automatically.

In a similar way one can get the coef\/f\/icients $C$ and $D$ from the equation
which can be obtained from (\ref{eqonAB}) by changing $A\to C$, $B\to D$
and $\Omega^{(1,3)}_{j,j'}\to \Omega^{(3,1)}_{j,j'}$ where $\Omega^{(3,1)}_{j,j'}$ is def\/ined through the equations (\ref{Omega}) and (\ref{eqonTheta}) with $I^{(+,1)}=\{3\}$, $I^{(-,1)}=\{1\}$.

The result for few leading orders with respect to $p$ looks as follows
\begin{gather}
 A=\frac1{4}  (p\nu)^{-1}+\frac{c - 22}{192}  (p\nu)^{-2}
-\left(\frac{\Delta(\Delta-1)}{16}+\frac{c^2+148c-860}{18432}\right)
  (p\nu)^{-3}
+O\big(p^{-4}\big),
\nonumber\\
 B=\frac18  (p\nu)^{-2}-\frac{c+14}{384}  (p\nu)^{-3}+O\big(p^{-4}\big).\label{A}
\end{gather}
The coef\/f\/icients $C$ and $D$ can be got from $A$ and $B$ respectively through
the substitution of $p$ by $-p$.

As in the $N=0$ case described above, one can substitute (\ref{trnasformM8}) into
(\ref{cft-6vexcited}) and use (\ref{basisoperator}) in order to get
 \begin{gather}
 A^2 \langle\Delta_-|L_2(U  V_{\mathbf{l}})\phi_{\al}(0)L_{-2}|\Delta_+\rangle
+B^2 \langle\Delta_-|L_1^2(U  V_{\mathbf{l}})\phi_{\al}(0)L_{-1}^2|\Delta_+\rangle \nonumber\\
\qquad {} +2AB \langle\Delta_-|L_1^2(U  V_{\mathbf{l}})\phi_{\al}(0)L_{-2}|\Delta_+\rangle
=V^{(1,3)},\label{eq13}
\end{gather}
where $V^{(1,3)}$ can be obtained from $V^{(0)}$ given by (\ref{V0}) via the substitution
$\Omega^{(0)}_{j,j'}\to \Omega^{(1,3)}_{j,j'}$
\begin{gather*}
V^{(1,3)}:=
\begin{pmatrix}
\big(\Omega^{(1,3)}_{1,7}+\Omega^{(1,3)}_{7,1}\big)/2\vspace{1mm}\\
(\Omega^{(1,3)}_{7,1}-\Omega^{(1,3)}_{1,7}\big)/(2d_\al)\vspace{1mm}\\
\big(\Omega^{(1,3)}_{3,5}+\Omega^{(1,3)}_{5,3}\big)/2\vspace{1mm}\\
\big(\Omega^{(1,3)}_{5,3}-\Omega^{(1,3)}_{3,5}\big)/(2d_\al)\vspace{1mm}\\
\Omega^{(1,3)}_{1,1}\Omega^{(1,3)}_{3,3}/3
-\Omega^{(1,3)}_{1,3}\Omega^{(1,3)}_{3,1}/4
\end{pmatrix}.
%\label{V13}
\end{gather*}
Both sides of the equation (\ref{eq13}) can be represented as a power series with respect to $p^{-1}$ starting with $p^0$. Then one can equate coef\/f\/icients standing at powers $p^{-j}$ and get as many equations as necessary. It turned out that all the equations obtained by taking coef\/f\/icients at~$p^0$ up to~$p^{-8}$ fulf\/ill automatically. The unknown matrix elements of the transformation matrix~$U$ are f\/ixed only in the order~$p^{-9}$. All equations which stem from further orders with respect to~$p^{-1}$ should be fulf\/illed automatically. Unfortunately, so far we could not prove it or even check any further equations because of complexity of calculations in the intermediate stage. We hope to do it in future.
We checked that if one substitutes $A\to C$, $B \to D$ in the left hand side of the equation~(\ref{eq13})  and $\Omega^{(1,3)}_{j,j'}\to \Omega^{(3,1)}_{j,j'}$ in the right hand side of~(\ref{eq13}) then the equation obtained is fulf\/illed automatically up to~$p^{-9}$.
Let us explicitly show the f\/inal result using even and odd combinations~(\ref{evenodd}):
\begin{gather}
 \phib_{1,7}^{\mathrm{even}}\cong\mathbf{l}_{-2}^4+
\bigl(4(c-4)\Delta+4(c+8)\bigr)/\bigl(3(\Delta+4)\bigr) \mathbf{l}_{-4}\mathbf{l}_{-2}^2\nn\\
\phantom{\phib_{1,7}^{\mathrm{even}}\cong}{}
 -\bigl(-43c^2+924c-16340-(21c^2+1102c-18535)\Delta\nn\\
 \phantom{\phib_{1,7}^{\mathrm{even}}\cong}{}
 -(11c^2-198c+775)\Delta^2+40(c-25)\Delta^3\bigr)/\bigl(45(\Delta+4)(\Delta+11)\bigr)
 \mathbf{l}_{-4}^2\nn\\
\phantom{\phib_{1,7}^{\mathrm{even}}\cong}{}
 -\bigl(2(-c^2-1540c+17264)-2(43c^2-298c+2652)\Delta\nn\\
 \phantom{\phib_{1,7}^{\mathrm{even}}\cong}{}
 -
12(c^2-31c+574)\Delta^2+32(c-28)\Delta^3\bigr)/\bigl(15(\Delta+4)(\Delta+11)\bigr)
 \mathbf{l}_{-6}\mathbf{l}_{-2}\nn\\
\phantom{\phib_{1,7}^{\mathrm{even}}\cong}{}
 -\bigl(-45c^3+1637c^2-137176c+2033360-20(2c^3+261c^2-7623c+67411)\Delta\nn\\
 \phantom{\phib_{1,7}^{\mathrm{even}}\cong}{}
 -4(10c^3-439c^2+7142c-5825)\Delta^2\nonumber\\
 \phantom{\phib_{1,7}^{\mathrm{even}}\cong}{}
 +96(2c^2-81c+705)\Delta^3\bigr)/\bigl(105
(\Delta+4)(\Delta+11)\bigr)%\nn\\
 \mathbf{l}_{-8},
\label{resM817}\\
 \phib_{1,7}^{\mathrm{odd}}\cong
\bigl(4\Delta\bigr)/(\Delta+4) \mathbf{l}_{-4}\mathbf{l}_{-2}^2\nn\\
\phantom{\phib_{1,7}^{\mathrm{odd}}\cong}{}
 +\bigl(20(c-28)+24(c+47)\Delta+8(2c-11)\Delta^2\bigr)/\bigl(15(\Delta+4)(\Delta+11)\bigr)
 \mathbf{l}_{-4}^2\nn\\
\phantom{\phib_{1,7}^{\mathrm{odd}}\cong}{}
 +\bigl(-40(c-28)+8(13c-44)\Delta+16(c-8)\Delta^2\bigr)/\bigl(5(\Delta+4)(\Delta+11)\bigr)
 \mathbf{l}_{-6}\mathbf{l}_{-2}\nn\\
\phantom{\phib_{1,7}^{\mathrm{odd}}\cong}{}
 -\bigl(20(c-28)(c+79)-4(15c^2+1316c-20143)\Delta\nn\\
 \phantom{\phib_{1,7}^{\mathrm{odd}}\cong}{}
 -4(15c^2-398c+4359)\Delta^2+
32(c-33)\Delta^3\bigr)/\bigl(35(\Delta+4)(\Delta+11)\bigr)
 \mathbf{l}_{-8},
\nn
\\
 \phib_{3,5}^{\mathrm{even}}\cong
 \mathbf{l}_{-2}^4+
\bigl(8(2c+13)+12(c-16)\Delta\bigr)/\bigl(9(\Delta+4)\bigr)
 \mathbf{l}_{-4}\mathbf{l}_{-2}^2\nn\\
\phantom{\phib_{3,5}^{\mathrm{even}}\cong}{}
 +\bigl(2(262c^2-4271c+82750)
+3(59c^2+2338c-67785)\Delta\nn\\
\phantom{\phib_{3,5}^{\mathrm{even}}\cong}{}
 +
(79c^2-1502c-6965)\Delta^2+120(c-25)\Delta^3\bigr)/\bigl(405(\Delta+4)(\Delta+11)\bigr)
 \mathbf{l}_{-4}^2\nn\\
\phantom{\phib_{3,5}^{\mathrm{even}}\cong}{}
 +\bigl(4(68c^2+7571c-86380)+4(188c^2-3379c+10310)\Delta
\nn\\
\phantom{\phib_{3,5}^{\mathrm{even}}\cong}{}
 +
24(4c^2-127c+1125)\Delta^2\bigr)/\bigl(135(\Delta+4)(\Delta+11)\bigr)
\mathbf{l}_{-6}\mathbf{l}_{-2}\nn\\
\phantom{\phib_{3,5}^{\mathrm{even}}\cong}{}
 +\bigl(2(420c^3-9013c^2+711929c-10449400)\nonumber\\
 \phantom{\phib_{3,5}^{\mathrm{even}}\cong}{}
 +4(70c^3+9741c^2-370938c+3325745)\Delta
\nn\\
\phantom{\phib_{3,5}^{\mathrm{even}}\cong}{}
 +20(14c^3-629c^2+9532c-48805)\Delta^2\nonumber\\
 \phantom{\phib_{3,5}^{\mathrm{even}}\cong}{}
 +96(c-25)(2c-31)\Delta^3\bigr)/
\bigl(945(\Delta+4)(\Delta+11)\bigr)%\nn\\
 \mathbf{l}_{-8},
\label{resM835}\\
 \phib_{3,5}^{\mathrm{odd}}\cong
\bigl(4\Delta\bigr)/\bigl(3(\Delta+4)\bigr) \mathbf{l}_{-4}\mathbf{l}_{-2}^2\nn\\
\phantom{\phib_{3,5}^{\mathrm{odd}}\cong}{}
 +\bigl(-4(17c-236)-24(c-43)\Delta+8(4c-127)\Delta^2\bigr)/\bigl(135(\Delta+4)(\Delta+11)\bigr)
 \mathbf{l}_{-4}^2\nn\\
\phantom{\phib_{3,5}^{\mathrm{odd}}\cong}{}
 +\bigl(8(17c-236)+8(47c-536)\Delta+48(c-18)\Delta^2\bigr)/\bigl(45(\Delta+4)(\Delta+11)\bigr)
 \mathbf{l}_{-6}\mathbf{l}_{-2}\nn\\
\phantom{\phib_{3,5}^{\mathrm{odd}}\cong}{}
 +\bigl(28(15c^2-571c+7708)+4(35c^2+2708c-66859)\Delta\nn\\
 \phantom{\phib_{3,5}^{\mathrm{odd}}\cong}{}
 +20(7c^2-222c+767)\Delta^2+
96(c-33)\Delta^3\bigr)/\bigl(315(\Delta+4)(\Delta+11)\bigr)
 \mathbf{l}_{-8}\nn
\end{gather}
and for the fourth-order combination we get
\begin{gather}
 \beta_1^{\rm CFT*}\beta_3^{\rm CFT*}\gamma_3^{\rm CFT*}\gamma_1^{\rm CFT*}\cong
1/12\cdot\mathbf{l}_{-2}^4+\bigl(3c-54+2(c-22)\Delta\bigr)/\bigl(18(\Delta+4)\bigr)
\mathbf{l}_{-4}\mathbf{l}_{-2}^2
\nn\\
\qquad{}
 -\bigl(
43c^2-1426c+11664+2(15c^2-302c-1069)\Delta\nn\\
\qquad{} +2(c^2+86c-2667)\Delta^2+
16(c-25)\Delta^3\bigr)/\bigl(216(\Delta+4)(\Delta+11)\bigr) \mathbf{l}_{-4}^2\nn\\
\qquad{}
 +\bigl(215c^2-8714c+74952+(125c^2-3856c+13208)\Delta\nn\\
\qquad{} +2(5c^2-73c-1816)\Delta^2+
16(c-28)\Delta^3\bigr)/\bigl(180(\Delta+4)(\Delta+11)\bigr) \mathbf{l}_{-6}\mathbf{l}_{-2}
\nn\\
\qquad{} -\bigl(-25c^3+1755c^2-39410c+326592-4(30c^2-1393c+9520)
\Delta\nn\\
\qquad{} +8(11c+289)\Delta^2+192\Delta^3\bigr)/\bigl(72(\Delta+4)(\Delta+11)\bigr) \mathbf{l}_{-8}.\label{resM8det}
\end{gather}
The elements of the above matrix $U$ can be easily got from these data.
It is interesting to note that the determinant of $U$ has relatively simple
factorized form
\begin{gather}
 \det U=
-32 (c-25) (c-28) (c-33) (\Delta-1)
\label{detU}\\
\phantom{\det U=}{} \times \frac{(c+2-2(c+11)\Delta+48\Delta^2)
(-c^2+18c+175+16(c-25)\Delta+192\Delta^2)}{382725(\Delta+4)(\Delta+11)}.
\nn
\end{gather}
The numerator of (\ref{detU}) is proportional to the following product
\begin{gather*}
 (\nu-2)^2(\nu-3)(\nu-4)(2\nu-3)(3\nu-4)(\al\nu-1)(\al\nu-2)
(\al\nu+1-\nu)(\al\nu-1-\nu)
%\label{numer}
\\
\qquad{}\times (\al\nu+2-\nu)(\al\nu-2+\nu)(\al\nu-2-\nu)
(\al\nu+1-2\nu)(\al\nu+2-2\nu)(\al\nu+2-3\nu).
\end{gather*}
It would be interesting to understand the meaning of the degeneration points
like $\nu=2$, $\nu=3$, $\nu=4$, $\nu=2/3$, $\nu=3/4$ and
$\al=1/\nu$, $\al=2/\nu$, $\al=-(1-\nu)/\nu$ etc.\ where the determinant
$\det U=0$.

\section{Conclusion}\label{section7}

In this paper we demonstrated that the method developed
in~\cite{HGSIV} works for excited states as well. As we saw with the
example of the descendant level $M=8$, it is not possible to
completely determine the transformation matrix between the usual
basis constructed through the action of Virasoro generators on the
primary f\/ield and the fermionic basis constructed
in~\cite{HGSI,HGSII} without involving excitations. On the other
hand the matrix elements of the transformation matrix should not
depend on the fact which excitation is taken. This should provide an
interesting compatibility condition on the structure of the
three-point functions both from the point of view of the CFT and the
lattice XXZ model we started with. In this paper we were able to
treat only the case of excitations corresponding to the descendant
level $N\le 2$. It would be interesting to check the above mentioned
compatibility for the case of other excitations with $N>2$ and for
the higher descendants $M>8$ as well. Of course, it is very
important to f\/ind a general proof of the compatibility condition.
We think it is also interesting to study singular points of the
transformation matrix mentioned in the end of the previous section.
This may shed new light on the structure of the Virasoro algebra,
Verma modules and singular vectors. One more important
generalization of the results obtained here, which is still out of
our reach, would be to treat the case of general values $\al$,
$\kappa$, $\kappa'$ and also the case of dif\/ferent excitations
inserted at $+\infty$ and $-\infty$ on the cylinder. In both cases
the function $\rho$ is not 1 and we would have to generalize the
whole Wiener--Hopf factorization technique. We hope to return to
these questions in future publications.

\appendix

\section[The function $\Phi^{(p)}$ and integrals of motion]{The function $\boldsymbol{\Phi^{(p)}}$ and integrals of motion}\label{AppendixA}

Here we show several further orders of the $1/p$-expansion of the functions used in Section~\ref{section3}
for the case of excitations with $k=1$. For simplicity we use shorter notation $m_0 \equiv I_1^{(+,1)}$, $m_1 \equiv I_1^{(-,1)}$.
First let us show expressions for several coef\/f\/icients $\Psi^{(p)}_j(l)$ of the expansion (\ref{expanPsi})
for the function $\Psi^{(p)}(l,p)$:
\begin{gather}
 \Psi^{(p)}_{0}(l) = -\frac{i}{l(l+i/(2\nu))},
\nn
\\
 \Psi^{(p)}_{1}(l) = 0,\nn\\
 \Psi^{(p)}_{2}(l) = \frac{i}{24}(-1+12(m_0+m_1)),\nn\\
 \Psi^{(p)}_{3}(l) = -\frac{l-i/(2\nu)}{8} (m_0^2-m_1^2),
\nn\\
 \Psi^{(p)}_{4}(l) = -\frac{l-i/(2\nu)}{2880\nu(1-\nu)}
\bigl(i l\nu(1-\nu)(7/2+60(m_0^3+m_1^3))+5(2\nu^2-11(1-\nu))(m_0^3+m_1^3)\nn\\
\phantom{\Psi^{(p)}_{4}(l) =}{}  +5(1+\nu)(2\nu-1)(m_0+m_1)(6(m_0+m_1)-1)+\nu^2+3(1-\nu)\bigr),\nn
\\
 \Psi^{(p)}_{5}(l) = \frac{l-i/(2\nu)}{147456\nu^2(1-\nu)^2} (m_0^2-m_1^2)
\Bigl(384 l^2\nu^2(1-\nu)^2(m_0^2+m_1^2)\nn
\\
\phantom{\Psi^{(p)}_{5}(l) =}{}  -16i l \nu(1-\nu)
\bigl(5(2\nu^2+11(1-\nu))(m_0^2+m_1^2)\nonumber\\
\phantom{\Psi^{(p)}_{5}(l) =}{}
+2(1+\nu)(2\nu-1)(12(m_0+m_1)-1)\bigr)\nn\\
\phantom{\Psi^{(p)}_{5}(l) =}{}
 -(20\nu^4-220\nu^3+681\nu^2-922\nu+461)(m_0^2+m_1^2)\nn\\
\phantom{\Psi^{(p)}_{5}(l) =}{}
 -2(1+\nu)(2\nu-1)(2\nu^2+23(1-\nu))(12(m_0+m_1)-1))\Bigr),
\nn
\\
 \Psi^{(p)}_{6}(l) = \frac{l-i/(2\nu)}{8360755200\nu^3(1-\nu)^3}
\Biggl\{8640 i l^3 (-31 + 252 (m_0^5 + m_1^5)) (1 - \nu)^3 \nu^3
\nn
\\
\phantom{\Psi^{(p)}_{6}(l) =}{}
 +1440 l^2 (1 - \nu)^2\nu^2
\Bigl(567 (2 \nu^2+11(1 - \nu))(m_0^5 + m_1^5)\nn\\
\phantom{\Psi^{(p)}_{6}(l) =}{}
 +21 (1 + \nu) (2 \nu -1) (m_0 + m_1) (7 - 10 (m_0^2+ m_1^2- m_0 m_1) +
120 (m_0^3+m_1^3))\nn\\
\phantom{\Psi^{(p)}_{6}(l) =}{}
- 164 \nu^2-755(1-\nu)\Bigr) -90 il \nu(1 - \nu)
\Bigl(21 (6085 - 12170 \nu + 8769 \nu^2 - 2684 \nu^3
\nn
\\
\phantom{\Psi^{(p)}_{6}(l) =}{}
+ 244 \nu^4)(m_0^5+m_1^5) +
(1+\nu)(2\nu-1)(420(34\nu^2+295(1-\nu))(m_0^4+m_1^4)\nonumber\\
\phantom{\Psi^{(p)}_{6}(l) =}{}
+840(14\nu^2+113(1-\nu))
(m_0^3m_1+m_0m_1^3)
\nn
\\
\phantom{\Psi^{(p)}_{6}(l) =}{}
 -2520(2\nu^2+23(1-\nu))m_0^2m_1^2+70(274\nu^2-257(1-\nu))(m_0^3+m_1^3)\nn\\
 \phantom{\Psi^{(p)}_{6}(l) =}{}
 +2520  (1 + \nu) (2 \nu-1 )(m_0 + m_1) (-m_0 - m_1 + 12 m_0 m_1)\nonumber\\
 \phantom{\Psi^{(p)}_{6}(l) =}{}
 +
7(158\nu^2+761(1-\nu))(m_0+m_1)) -96(159 - 318 \nu + 240 \nu^2 - 81 \nu^3 + 8 \nu^4)\Bigr)
\nn
\\
\phantom{\Psi^{(p)}_{6}(l) =}{}
 -21 (210403 - 631209 \nu + 813759 \nu^2 - 575503 \nu^3 + 219354 \nu^4 -
   36804 \nu^5 \nonumber\\
   \phantom{\Psi^{(p)}_{6}(l) =}{}
 + 1288 \nu^6)(m_0^5+m_1^5) -(1+\nu)(2\nu-1)\Bigl(420 (20281 - 40562 \nu + 25053 \nu^2 \nonumber\\
 \phantom{\Psi^{(p)}_{6}(l) =}{}
 - 4772 \nu^3 + 52 \nu^4)
(m_0^4+m_1^4)
 -840 (-6407 + 12814 \nu - 6771 \nu^2 + 364 \nu^3 \nonumber\\
 \phantom{\Psi^{(p)}_{6}(l) =}{}
 + 436 \nu^4)m_0m_1(m_0^2+m_1^2)
 -2520 (2489 - 4978 \nu + 3837 \nu^2 - 1348 \nu^3 \nonumber\\
\phantom{\Psi^{(p)}_{6}(l) =}{}
+ 308 \nu^4)m_0^2m_1^2
 +70 (-31607 + 63214 \nu + 16989 \nu^2 - 48596 \nu^3\nn\\
 \phantom{\Psi^{(p)}_{6}(l) =}{}
+ 3316 \nu^4)(m_0^3+m_1^3) +12600  (1 + \nu) (-1 + 2 \nu) (2 \nu^2+35(1-\nu) ) (m_0 + m_1)\nonumber\\
\phantom{\Psi^{(p)}_{6}(l) =}{}
\times (-m_0 - m_1 + 12 m_0 m_1)
 -7 (-39599 + 79198 \nu - 52347 \nu^2 + 12748 \nu^3\nonumber\\
 \phantom{\Psi^{(p)}_{6}(l) =}{}
  + 2452 \nu^4)(m_0+m_1)\Bigr)
 +103680  (1 - \nu) (5 - 10 \nu + 10 \nu^2 - 5 \nu^3 + \nu^4)\Biggr\}.
\label{Psi-1to5}
\end{gather}
One can check that the $\Psi$-function in the ground state case is reproduced if one
takes $m_0=m_1=0$. There is a connection of the function $\Psi$ with the integrals
of motion described in \cite{BLZII}. In the ground state case this connection
was given by the formula (10.17) of \cite{HGSIV}:
\begin{gather}
 I_{2n-1}=-i\Psi^{(p)}\left(\frac{i(2n-1)}{2\nu},p\right)
 n(2n-1)(2\nu^2)^{n-1}p^{2n-1}.
\label{I2n-1}
\end{gather}
In the case of excitations the integrals of motion are given by
matrices. Let us consider the f\/irst three integrals of motion
$I_1$, $I_3$, $I_5$. Their explicit expressions via the Virasoro
generators can be found in~\cite{BLZI}.

Again let us consider an example of the excitations with $N=2$ and $k=1$ with two
possibilities $I^{(+,1)}=\{1\}$, $I^{(-,1)}=\{3\}$ and $I^{(+,1)}=\{3\}$, $I^{(-,1)}=\{1\}$ which correspond
to the two-dimensional descendant space spanned by two vectors
$L_{-2}|\Delta_+\rangle$ and $L_{-1}^2|\Delta_+\rangle$ or their linear combinations
$|\Delta_+ ;\{1\},\{3\}\rangle$, $|\Delta_+ ;\{3\},\{1\}\rangle$ given by the formula (\ref{13}).

Let us start with the simplest case of the very f\/irst integral of motion $I_1$ given by
the formula~(11) of the paper~\cite{BLZI}:
\begin{gather*}
I_1=L_0-\frac{c}{24}.
\end{gather*}
This operator is diagonal for the above two-dimensional space
\begin{gather*}
I_1
\begin{pmatrix}
L_{-2}|\Delta_+\rangle\\
L_{-1}^2|\Delta_+\rangle
\end{pmatrix}
=\left(\Delta_++2-\frac{c}{24}\right)
\begin{pmatrix}
L_{-2}|\Delta_+\rangle\\
L_{-1}^2|\Delta_+\rangle
\end{pmatrix}
\end{gather*}
and
\begin{gather*}
\Delta_++2-\frac{c}{24} = 2(p\nu)^2+\frac{47}{24}.
\end{gather*}
We can easily check that we get exactly the same result if we substitute the expansion
(\ref{expanPsi}) with the coef\/f\/icients given by (\ref{Psi-1to5}) for $l=i/(2\nu)$
and $m_0=1$, $m_1=3$ or $m_0=3$, $m_1=1$
into the formula (\ref{I2n-1}) with $n=1$. We see that in this case all
$\Psi^{(p)}_{j}(i/(2\nu))=0$ with $j\ge 3$ even without f\/ixing $m_0$ and $m_1$.

The next case of $I_3$ is a bit less trivial since $I_3$ is not diagonal any longer.
Following \cite{BLZI}, we have
\begin{gather*}
I_3 = 2\sum_{n=1}^{\infty} L_{-n}L_n + L_0^2 -\frac{c+2}{12} L_0 + \frac{c(5c+22)}{2880}.
\end{gather*}
We see that only the f\/irst term here is not diagonal.
Using the Virasoro algebra, one can check that
\begin{gather*}
 I_3
\begin{pmatrix}
L_{-2}|\Delta_+\rangle\\
L_{-1}^2|\Delta_+\rangle
\end{pmatrix}
=\begin{pmatrix}
c & 6\\
12\Delta_+ & 4
\end{pmatrix}
\begin{pmatrix}
L_{-2}|\Delta_+\rangle\\
L_{-1}^2|\Delta_+\rangle
\end{pmatrix}%\label{I3N2}
\\
 \hphantom{I_3
\begin{pmatrix}
L_{-2}|\Delta_+\rangle\\
L_{-1}^2|\Delta_+\rangle
\end{pmatrix}=}{} +\left(8\Delta_+
+(\Delta_+ +2)^2 -\frac{(c+2)(\Delta_+ +2)}{12} + \frac{c(5c+22)}{2880}\right)
 \begin{pmatrix}
L_{-2}|\Delta_+\rangle\\
L_{-1}^2|\Delta_+\rangle
\end{pmatrix}.
\nn
\end{gather*}
We can diagonalize the matrix
\begin{gather*}
\begin{pmatrix}
c & 6\\
12\Delta_+ & 4
\end{pmatrix}
=
{\begin{pmatrix}
C & D\\
A & B
\end{pmatrix}}^{-1}
\begin{pmatrix}
\la^+ & 0\\
0 & \la^-
\end{pmatrix}
\begin{pmatrix}
C & D\\
A & B
\end{pmatrix}
\end{gather*}
with the eigenvalues
\begin{gather}
\la^{\pm} = -\frac{6\nu^2-5(1-\nu)}{2(1-\nu)}
\pm \frac{3\sqrt{64\nu^2(1-\nu)^2p^2
+(1+\nu)^2(2\nu-1)^2}}{2(1-\nu)}
\label{la}
\end{gather}
and the matrix elements $A$, $B$, $C$, $D$ taken from the def\/inition
of the vectors $|\Delta_+ ;\{1\},\{3\}\rangle$,
$|\Delta_+ ;\{3\},\{1\}\rangle$ given by (\ref{13}). Hence these vectors
are eigenvectors of the matrix $I_3$.
Several leading terms of $1/p$-expansion of $A$, $B$ where shown in (\ref{A}) while
$C$, $D$ can be got from $A$, $B$ by changing $p\to -p$.
So, for the eigenvalues of $I_3$ we obtain\footnote{We hope the reader will not mix these eigenvalues $I_j^{(\pm)}$ with
the integrals of motion $I_j^{\pm}$ mentioned  in Section~\ref{section2} that equal to each other
when $\kappa'=\kappa$ and $\Delta_+=\Delta_-$.}
\begin{gather}
 I_3^{(\pm)} = \la^{\pm} +8\Delta_+
+(\Delta_+ +2)^2 -\frac{(c+2)(\Delta_+ +2)}{12} + \frac{c(5c+22)}{2880}
\nn\\
 \phantom{I_3^{(\pm)}}{}
 \simeq 4(p\nu)^4+\frac{47}2(p\nu)^2\mp 12p\nu-
\frac{4804\nu^2-5769(1-\nu)}{960(1-\nu)}\nonumber\\
 \phantom{I_3^{(\pm)} =}{}
\mp\frac{3}{32}
\frac{(1+\nu)^2(2\nu-1)^2}{(1-\nu)^2}(p\nu)^{-1}+O\big(p^{-2}\big).
\label{expanI3}
\end{gather}
We can check at least for number of f\/irst orders
that we get the same $1/p$-expansion if, like in the previous case
we substitute the expansion (\ref{expanPsi}) this time for $l=3i/(2\nu)$
and $m_0=3$, $m_1=1$ in case of $I^{(+)}_3$ and $m_0=1$, $m_1=3$ in case
of $I^{(-)}_3$  into the formula (\ref{I2n-1}) with $n=2$.
In contrast to the case of the f\/irst integral of motion~$I_1$,
the series expansion~(\ref{expanI3}) does not terminate because of the square root
in the expression~(\ref{la}) for the eigenvalues $\la^{\pm}$. Also we get
both even and odd powers with respect to~$p$ in contrast to the case $N=0$
corresponding to the ground state.

The next case can be treated similarly. First we take~$I_5$ from \cite{BLZI}
\begin{gather*}
 I_5=\sum\limits_{n_1+n_2+n_3=0} :L_{n_1}L_{n_2}L_{n_3}:+
\sum_{n=1}^{\infty}\left(\frac{c+11}6 n^2-1-\frac{c}4\right)L_{-n}L_n\nn\\
\phantom{I_5=}{}  +\frac32\sum_{r=1}^{\infty} L_{1-2r}L_{2r-1}-\frac{c+4}8L_0^2
+\frac{(c+2)(3c+20)}{576}L_0-\frac{c(3c+14)(7c+68)}{290304}
\end{gather*}
with the normal ordering : : for which the Virasoro generators with
bigger indices are placed to the right.
Then using the Virasoro algebra, one can come to the following formula
\begin{gather*}
I_5
\begin{pmatrix}
L_{-2}|\Delta_+\rangle\\
L_{-1}^2|\Delta_+\rangle
\end{pmatrix}
=5\left(\frac56+\frac{c}{24}+\Delta_+\right)
\begin{pmatrix}
c & 6\\
12\Delta_+ & 4
\end{pmatrix}
\begin{pmatrix}
L_{-2}|\Delta_+\rangle\\
L_{-1}^2|\Delta_+\rangle
\end{pmatrix}\\ %\label{I5N2}\\
\hphantom{I_5
\begin{pmatrix}
L_{-2}|\Delta_+\rangle\\
L_{-1}^2|\Delta_+\rangle
\end{pmatrix}=}{}
+\left(\Delta_+^3+\frac{236-c}8\Delta_+^2+
\left(\frac{c^2}{192}-\frac{227c}{288}+\frac{3845}{72}\right)\Delta_+\right)\nn\\
\hphantom{I_5
\begin{pmatrix}
L_{-2}|\Delta_+\rangle\\
L_{-1}^2|\Delta_+\rangle
\end{pmatrix}=}{}
+\left(\Delta_+^3
-\frac{c^3}{13824}+\frac{1361c^2}{145152}-\frac{7325c}{5184}
+\frac{221}{36}
\right)
 \begin{pmatrix}
L_{-2}|\Delta_+\rangle\\
L_{-1}^2|\Delta_+\rangle
\end{pmatrix}.
\nn
\end{gather*}
We see that the matrix $I_5$ can be diagonalized  by the same similarity transformation
as $I_3$ because all the integrals of motion commute.
The result for the asymptotic expansion of the two corresponding
eigenvalues $I_5^{\pm}$ looks
\begin{gather*}
I_5^{(\pm)}\simeq 8(p\nu)^6+\frac{235}2(p\nu)^4\mp 120(p\nu)^3-
\frac{4808\nu^2-12503(1-\nu)}{96(1-\nu)}(p\nu)^2
\nn\\
\phantom{I_5^{(\pm)}\simeq}{}
\mp\frac{15(4\nu^4+36\nu^3+21\nu^2-114\nu+57)}{16(1-\nu)^2}(p\nu)\nn\\
\phantom{I_5^{(\pm)}\simeq}{}
+\frac{822432\nu^4+2338220\nu^3-821605\nu^2-3033230\nu+1516615}{96768(1-\nu)^2}
+O\big(p^{-1}\big).
\end{gather*}
Again this result perfectly matches the expansion of (\ref{I2n-1}) for $n=3$ with
$\Psi^{(p)}\left(5i/(2\nu),p\right)$ taken for $m_0=3$, $m_1=1$ in case of $I_5^{(+)}$ and
for $m_0=1$, $m_1=3$ in case of $I_5^{(-)}$. Similar to $I_3^{(\pm)}$, the integral $I_5^{(\pm)}$ and all further integrals $I_{2n-1}^{(\pm)}$ are not polynomials with respect to~$p$ in contrast to the ground state case.

\section[The functions $\bar{F}(x,p)$ and $x_r^{\pm}(p)$]{The functions $\boldsymbol{\bar{F}(x,p)}$ and $\boldsymbol{x_r^{\pm}(p)}$}\label{AppendixB}

In Section~\ref{section3} we discussed several functions def\/ined within the TBA approach. In the previous appendix we showed few orders of the function $\Psi^{(p)}(l,p)$.  We also need the functions
$\bar{F}(x,p)$ and $x_r^{\pm}(p)$.  As was described in Section~\ref{section3}, the asymptotic expansions of these functions and the function $\Psi^{(p)}(l,p)$ are calculated order by order with respect to $1/p$ via the iterative procedure.
There we explained how does the very f\/irst iteration work.
Once the function $\Psi^{(p)}(l,p)$ is found up to some order, the next order of the function $\bar{F}(x,p)$ can be obtained via the equation~(\ref{TayF}). The coef\/f\/icients of (\ref{expanx}) are determined from the equation
(\ref{barF}). The further iteration steps are straightforward but the answer becomes rather cumbersome already after several iterations. For the reader who wants to check his own calculations we show few orders of the expansions~(\ref{expanF}) and~(\ref{expanx}) in the case $k=1$. We do not think it would be instructive to show further orders. As in the previous appendix we use the shorthand notation $m_0 \equiv I_1^{(+,1)}$, $m_1 \equiv I_1^{(-,1)}$.
\begin{gather*}
 \bar{F}_0(x)=-i\frac{(1+\nu)(2\nu-1)}{576\nu(1-\nu)}(1-12(m_0+m_1))-
ix^2\frac{2\nu^2+11(1-\nu)}{48\nu(1-\nu)},
\nn
\\
 \bar{F}_1(x)=i\frac{(1+\nu)(2\nu-1)(2\nu^2+23(1-\nu))}{9216\nu^2(1-\nu)^2}-
x\frac{(1+\nu)^2(2\nu-1)^2}{27648\nu^2(1-\nu)^2}(1-12(m_0+m_1))
\nn\\
\phantom{\bar{F}_1(x)=}{}
+x^3\frac{4\nu^4-44\nu^3+309\nu^2-530\nu+265}{6912\nu^2(1-\nu)^2},
\nn\\
 \bar{F}_2(x)=-\frac{i(1+\nu)(2\nu-1)}{2388787200\nu^3 (1 - \nu)^3 }
\Bigl(120(436\nu^4+364\nu^3-6771\nu^2+12814\nu-6407)\nn\\
\phantom{\bar{F}_2(x)=}{}
\times (m_0^3+m_1^3) +3600(1+\nu)(2\nu-1)(2\nu^2+23(1-\nu))(m_0+m_1)\nn\\
\phantom{\bar{F}_2(x)=}{}
 \times  (1-6(m_0+m_1)) +2452\nu^4+9148\nu^3-50547\nu^2+82798\nu-41399\Bigr)\nn\\
\phantom{\bar{F}_2(x)=}{}
+\frac{x(1+\nu)(2\nu-1)}{3317760 \nu^3(1 - \nu)^3 }
(556\nu^4-1676\nu^3+2859\nu^2-2366\nu+1183)(m_0^2-m_1^2)\nn\\
\phantom{\bar{F}_2(x)=}{}
 -\frac{ix^2(1+\nu)(2\nu-1)}{19906560 \nu^3(1 - \nu)^3 }
(556\nu^4-2036\nu^3+3039\nu^2-2006\nu+1003)\nn\\
\phantom{\bar{F}_2(x)=}{}
\times (1-12(m_0+m_1)) -\frac{ix^4}{9953280  \nu^3(1 - \nu)^3 }
(1112\nu^6-2796\nu^5-5154\nu^4\\
\phantom{\bar{F}_2(x)=}{}
+64603\nu^3-154059\nu^2+146109\nu-48703),
\nn
\\
 \bar{F}_3(x)=-i\frac{(1+\nu)(2\nu-1)}{15288238080 \nu^4 (1 - \nu)^4}
(m_0^2-m_1^2)\Bigl(3(7736\nu^6-49644\nu^5+41430\nu^4\nn\\
\phantom{\bar{F}_3(x)=}{}
 +233059\nu^3
-658107\nu^2+649893\nu-216631)(m_0^2+m_1^2) \nn\\
\phantom{\bar{F}_3(x)=}{}
 -4(1+\nu)(2\nu-1)(932\nu^4+788\nu^3-22227\nu^2+42878\nu-21439)
\nn\\
\phantom{\bar{F}_3(x)=}{} \times
(1-12(m_0+m_1))\Bigr)
 +x\frac{(1+\nu)(2\nu-1)}{229323571200  \nu^4 (1 - \nu)^4}
\Bigl(1080(1496\nu^6-12924\nu^5\\
\phantom{\bar{F}_3(x)=}{}
+42270\nu^4-75161\nu^3+78753\nu^2  -49407\nu+16469)(m_0^3+m_1^3)\nn\\
\phantom{\bar{F}_3(x)=}{}
 -960 (1+\nu)(2\nu-1)(556\nu^4-1676\nu^3+2859\nu^2-2366\nu+1183)
(m_0+m_1)\\
\phantom{\bar{F}_3(x)=}{}\times
(1-6(m_0+m_1))  +138728\nu^6-926052\nu^5+2802450\nu^4-
4743023\nu^3\\
\phantom{\bar{F}_3(x)=}{}
+4847079\nu^2-2970681\nu+990227\Bigr)  +i x^2\frac{(1+\nu)(2\nu-1)}{637009920  \nu^4 (1 - \nu)^4}
(m_0^2-m_1^2)\\
\phantom{\bar{F}_3(x)=}{}
\times (4568\nu^6-40572\nu^5+138270\nu^4-239513\nu^3
+230049\nu^2
 -132351\nu+44117)\\
\phantom{\bar{F}_3(x)=}{}
+x^3\frac{(1+\nu)^2(2\nu-1)^2}{5733089280  \nu^4 (1 - \nu)^4}
(2284\nu^4-8084\nu^3+12111\nu^2 -8054\nu+4027)\nn\\
\phantom{\bar{F}_3(x)=}{} \times
(1-12(m_0+m_1))+
\frac{x^5}{4777574400 \nu^4 (1 - \nu)^4}
(9136\nu^8-76576\nu^7+196840\nu^6\nn\\
\phantom{\bar{F}_3(x)=}{}
 +149096\nu^5-3382325\nu^4+
10503740\nu^3-14391926\nu^2+9334868\nu-2333717),
\nn
\\
 \bar{F}_4(x)=\frac{i(1+\nu)(2\nu-1)}{2157476157849600
\nu^5(1 - \nu)^5}\Bigl(
6048(4048\nu^8+208064\nu^7-1145688\nu^6+1757040\nu^5\!\nn\\
\phantom{\bar{F}_4(x)=}{}
 +1432457\nu^4-8416452\nu^3+12210950\nu^2-8061828\nu+
2015457)(m_0^5+m_1^5)\nn\\
\phantom{\bar{F}_4(x)=}{}
 +3809872 \nu^8-215990816 \nu^7
+939983096 \nu^6-1142084632 \nu^5-1598906179 \nu^4\\
\phantom{\bar{F}_4(x)=}{}
+6538494484 \nu^3
 -8737091978 \nu^2+5620794700 \nu -1405198675\nn\\
\phantom{\bar{F}_4(x)=}{}
 -196(1+\nu)(2\nu-1)(m_0+m_1) \bigl(
240(11528\nu^6-109308\nu^5+124842\nu^4+705307\nu^3\\
\phantom{\bar{F}_4(x)=}{}
-2193591\nu^2+2209125\nu-736375)(m_0^3+m_1^3)
 +120(27112\nu^6-287566\nu^5\\
\phantom{\bar{F}_4(x)=}{}
 +627666\nu^4+187193\nu^3
-2262129\nu^2+2602239\nu -867413)m_0m_1(m_0^2+m_1^2)
\nn\\
\phantom{\bar{F}_4(x)=}{}
+10(247064\nu^6+795588\nu^5-9542418\nu^4+12330151\nu^3+
6743697\nu^2 -15490527\nu\nn\\
\phantom{\bar{F}_4(x)=}{}
+5163509)(m_0^2+m_1^2)
 +10(575464\nu^6+728508\nu^5-17201838\nu^4+25221881\nu^3\\
 \phantom{\bar{F}_4(x)=}{}
+6701007\nu^2-23174337\nu+7724779)m_0m_1
 +360(2\nu^2+23(1-\nu))(676\nu^4\\
 \phantom{\bar{F}_4(x)=}{}
 -3716\nu^3+12489\nu^2
-17546\nu+8773)m_0m_1(m_0+m_1) -5040(1+\nu)(2\nu-1)\nn\\
 \phantom{\bar{F}_4(x)=}{} \times
(68\nu^4+92\nu^3-2223\nu^2+4262\nu-2131)
(m_0+m_1) +7(35272\nu^6\nn\\
 \phantom{\bar{F}_4(x)=}{}
-272436\nu^5+362346\nu^4+559733\nu^3
-2128749\nu^2+2218659\nu-739553)\bigr)\Bigr)\nn\\
\phantom{\bar{F}_4(x)=}{}
 -\frac{x(1+\nu)(2\nu-1)}{1284211998720 \nu^5(1 - \nu)^5}(m_0^2-m_1^2)
\Bigl(8 ( 114704 \nu^8- 186848 \nu^7- 1703896 \nu^6 \nn\\
\phantom{\bar{F}_4(x)=}{}
+ 8391976 \nu^5
 - 17873683 \nu^4  + 22437364 \nu^3  - 17519338 \nu^2 + 8605900 \nu
   -2151475)\nn\\
\phantom{\bar{F}_4(x)=}{}\times
 (m_0^2+m_1^2)   + 7  (1 + \nu) (2 \nu-1 ) ( 11240 \nu^6
   - 137412 \nu^5 + 480594 \nu^4  - 893735 \nu^3 \\
\phantom{\bar{F}_4(x)=}{}
   +
   965295 \nu^2 - 622113 \nu
     + 207371)(1 - 12(m_0 + m_1))\Bigr)\\
\phantom{\bar{F}_4(x)=}{}
-i\frac{x^2(1+\nu)^2(2\nu-1)^2}{77052719923200 \nu^5(1 - \nu)^5}
\Bigl(120 ( 1247080 \nu^6 - 6770580 \nu^5 +
   15390258 \nu^4\nn\\
\phantom{\bar{F}_4(x)=}{}
 - 18566683 \nu^3 + 12601659 \nu^2 - 3981981 \nu
+1327327)(m_0^3+m_1^3)\nn\\
\phantom{\bar{F}_4(x)=}{}
 + 1680  ( 4568 \nu^6  - 40572 \nu^5 + 138270 \nu^4 -
   239513 \nu^3 + 230049 \nu^2 - 132351 \nu\nn\\
\phantom{\bar{F}_4(x)=}{}
     + 44117)(m_0 + m_1) (1 - 6( m_0 + m_1))
+7 ( 1201400 \nu^6 - 6364860 \nu^5  +
   14007558 \nu^4  \nn\\
\phantom{\bar{F}_4(x)=}{}
 - 16171553 \nu^3   + 10301169 \nu^2 - 2658471 \nu + 886157)\Bigr)\nn\\
 \phantom{\bar{F}_4(x)=}{}
 + \frac{x^3(1+\nu)(2\nu-1)}{321052999680\nu^5(1 - \nu)^5} (m_0^2-m_1^2)
( 2622064 \nu^8 - 14133568 \nu^7 + 33585976 \nu^6\nn\\
 \phantom{\bar{F}_4(x)=}{}
 - 49220560 \nu^5\!  +  55866331 \nu^4\! - 52813996 \nu^3 \! + 43115890 \nu^2\! - 21866764 \nu+
5466691) \nn\\
 \phantom{\bar{F}_4(x)=}{}
 - \frac{i x^4(1+\nu)(2\nu-1)}{3852635996160\nu^5(1 - \nu)^5}
(2622064 \nu^8 - 14517280 \nu^7 + 35135944 \nu^6 \\
 \phantom{\bar{F}_4(x)=}{}
 - 51550552 \nu^5\!
 +  56686759 \nu^4 \!
   - 51453364 \nu^3\! + 41083762 \nu^2\! - 20513692 \nu
+5128423)\\
\phantom{\bar{F}_4(x)=}{}
 \times
(1-12(m_0+m_1))
 - \frac{i x^6}{4815794995200\nu^5(1 - \nu)^5}
  (5244128 \nu^10 - 25645072 \nu^9\\
 \phantom{\bar{F}_4(x)=}{}
   +
  45932736 \nu^8 - 30480936 \nu^7 + 22676526 \nu^6
    - 291308781 \nu^5
   +  1140456861 \nu^4\\
 \phantom{\bar{F}_4(x)=}{}
    - 2039726202 \nu^3 + 1941710088 \nu^2 - 954519025 \nu
+  190903805).\nn
\end{gather*}
For the parameters $x^{\pm}_1(p)\equiv x^{\pm}(p) =
\sum\limits_{j=0}^{\infty}x^{\pm}_j p^{-j}$ which determine the Bethe roots
via the relation~(\ref{expanx}) we had the initial conditions $x^+_0=\frac{im_0}2$,
$x^-_0=-\frac{im_1}2$ and then
\begin{gather*}
 x^+_1=\frac{i(2\nu^2+11(1-\nu))}{192\nu(1-\nu)}m_0^2
-\frac{i(1+\nu)(2\nu-1)}{576\nu(1-\nu)}(1-12(m_0+m_1)),
\nn\\
x^+_2=\frac{i(20 \nu^4 - 220 \nu^3 + 681 \nu^2 - 922 \nu +461)
m_0^3}{55296 \nu^2 (1 - \nu)^2}
\nn\\
\phantom{x^+_2=}{}
+\frac{i (1+\nu)(2\nu-1)(2\nu^2+23(1-\nu))}{55296 \nu^2 (1 - \nu)^2}
(-m_0+18m_0^2+12m_0m_1-6m_1^2),\nn\\
\nn\\
x^+_3=
\frac{i (1288 \nu^6 - 36804 \nu^5 + 219354 \nu^4 - 575503 \nu^3 + 813759 \nu^2 - 631209 \nu + 210403)m_0^4}{159252480 \nu^3(1-\nu )^3}\nn\\
\phantom{x^+_3=}{}
+\frac{i (1 + \nu) (2 \nu-1)
(52 \nu^4 - 4772 \nu^3 + 25053 \nu^2 - 40562 \nu + 20281)m_0^3}{9953280 \nu^3(1 - \nu)^3}\nn\\
\phantom{x^+_3=}{}
-\frac{i  (1 + \nu) (2\nu-1 )
(436 \nu^4 + 364 \nu^3 - 6771 \nu^2 + 12814 \nu -6407)m_1
(3 m_0^2 + m_1^2)}{19906560 \nu^3 (1 - \nu)^3 }\nn\\
\phantom{x^+_3=}{}
+\frac{i (1 + \nu)^2 (2 \nu -1)^2 (2 \nu^2 +35(1 - \nu) )
(-m_0 - m_1 + 12 m_0 m_1 + 6 m_1^2)}{663552 \nu^3(1 - \nu)^3}\nn\\
\phantom{x^+_3=}{}
+\frac{i  (1 + \nu) (2 \nu-1 ) (3316 \nu^4 - 48596 \nu^3 + 16989 \nu^2 + 63214 \nu -31607)
m_0^2}{79626240 \nu^3 (1 - \nu)^3}\nn\\
\phantom{x^+_3=}{}
-\frac{i  (1 + \nu) (2 \nu-1 ) (308 \nu^4 -
   1348 \nu^3 + 3837 \nu^2 - 4978 \nu +2489)m_0 m_1^2}{3317760 \nu^3(1 - \nu)^3}\nn\\
\phantom{x^+_3=}{}
-\frac{i (1 + \nu) (2 \nu-1) (2452 \nu^4 +
     12748 \nu^3 - 52347 \nu^2 + 79198 \nu-39599)}{2388787200 \nu^3 (1 - \nu)^3},
\nn\\
x^+_4=\frac{i }{458647142400 \nu^4(1 - \nu)^4}
\Bigl(-3\bigl(18704 \nu^8+ 1330336 \nu^7 - 16442440 \nu^6 + 75087544 \nu^5\nn\\
\phantom{x^+_4=}{}
-183046975 \nu^4  + 267969460 \nu^3- 244932514 \nu^2+ 133379452 \nu
-33344863\bigr)m_0^5\nn\\
\phantom{x^+_4=}{}
-150 (1 + \nu) (2 \nu-1 )
\bigl(3256 \nu^6 + 51156 \nu^5 - 738570 \nu^4 + 2884979 \nu^3
 - 5217867 \nu^2\nn\\
\phantom{x^+_4=}{}
 + 4530453 \nu-1510151\bigr)m_0^4
 +30 (1 + \nu) (2 \nu-1 )
\bigl(53816 \nu^6 - 1086444 \nu^5+ 6642390 \nu^4\\
\phantom{x^+_4=}{}
 - 7848221 \nu^3
 - 4235067 \nu^2  + 9791013 \nu -3263671\bigr)m_0^3
-90  (1 + \nu) (2 \nu-1 )
\bigl(7736 \nu^6\\
\phantom{x^+_4=}{}
- 49644 \nu^5 + 41430 \nu^4
+ 233059 \nu^3 - 658107 \nu^2
 + 649893 \nu -216631\bigr)m_1 (4 m_0^3 - m_1^3)
\nn\\
\phantom{x^+_4=}{}
+7200 (1 + \nu)^2 ( 2 \nu-1)^2 (2 \nu^2+23(1 - \nu) )^2
(-3 m_0^2 - 2 m_0 m_1 + 36 m_0^2 m_1 + m_1^2\nn\\
\phantom{x^+_4=}{}
 - 12 m_1^3)
+  (1 + \nu) (2 \nu-1)
\bigl(74632 \nu^6 - 808308 \nu^5 + 3707610 \nu^4 -8492107 \nu^3\\
\phantom{x^+_4=}{}
+ 10979811 \nu^2 - 8080509 \nu
 + 2693503\bigr)m_0-60 (1 + \nu) (2 \nu-1)
\bigl(9976 \nu^6 - 100044 \nu^5\\
\phantom{x^+_4=}{}
 + 431430 \nu^4 - 1092901 \nu^3 +1621773 \nu^2
- 1290387 \nu + 430129\bigr)
m_0 m_1^2 (3 m_0 - 2 m_1)\Bigr).
\nn
\end{gather*}
The coef\/f\/icients $x^-_j$ can be got from $\tilde x^+_j$ obtained from $x^+_j$ by the replacement
$m_0\leftrightarrow m_1$ as follows:
\begin{gather*}
x^-_{2j}=-\tilde{x}^+_{2j},\qquad x^-_{2j+1}=\tilde{x}^+_{2j+1}, \qquad j=0,1,\dots.
\end{gather*}

\section[The function $\omegab$]{The function $\boldsymbol{\omegab}$}\label{AppendixC}

In Section~\ref{section4} we explained how we determine our main object: the function $\omegab$. The asymptotic expansion
for $\omegab$ is related to the function $\Theta$ by the formula (\ref{asym-omega}). In it's turn the coef\/f\/icients~$\Theta_n$ of asymptotic expansion (\ref{asyTheta}) of the function $\Theta$
\begin{gather*}
\Theta(l,m|p,\al)
\simeq
\sum_{j=0}^\infty \Theta_j(l,m|\al)p^{-j}
\end{gather*}
can be found with help of the TBA data of Section~\ref{section3} and Appendices~\ref{AppendixA},~\ref{AppendixB}
via solving the equation~(\ref{eqonTheta}) by iterations.
In this appendix we show the result for several leading coef\/f\/icients again in the case $k=1$.
With the exception of the coef\/f\/icient~$\Theta_0$, all other coef\/f\/icients $\Theta_j$ are polynomials
with respect to~$l$ and~$m$:
\begin{gather*}
\Theta_0(l,m|\al)=-\frac{i}{l+m},\\ %\label{Theta0to4}\\
\Theta_1(l,m|\al)=0,\nn\\
\Theta_2(l,m|\al)=-\frac{\nu^2 \al(2-\al)+2i(1-\nu)(-i+2\nu(l+m))}
{96\nu(1-\nu)} (1-12(m_0+m_1)),
\nn\\
\Theta_3(l,m|\al)=-\frac{m_0^2-m_1^2}{512\nu^2(1-\nu)^2}
 \Bigl(-3\nu^4\al^2(2-\al)^2-24i \al(2-\al)(1-\nu)\nu^2(-i+\nu(l+m))
\nn\\
\phantom{\Theta_3(l,m|\al)=}{}
+32(1-\nu)^2(-i+\nu(l+m))\;(-i+2\nu(l+m))
\Bigr),
\\
\Theta_4(l,m|\al)=-\frac{1}{184320\nu^3 (1 - \nu)^3}
\Bigl(-5\nu^6
\bigl(5(m_0^3+m_1^3)-42(m_0+m_1)^2+7(m_0+m_1)\bigr)\\
\phantom{\Theta_4(l,m|\al)=}{}
\times \al^3(2-\al)^3\nn
+10 i \nu^4(1 - \nu)\bigl(-24  (-3i+2(l+m)\nu)(m_0^3+m_1^3)\\
\phantom{\Theta_4(l,m|\al)=}{}
+2 m^{(0,1)} (-16 i+9(l+m)\nu)
-(-i + (l + m)\nu)\bigr) \al^2(2-\al)^2
+8 \nu(1-\nu)\\
\phantom{\Theta_4(l,m|\al)=}{}\times
\bigl(2\nu(1 - \nu)  (7 + 120 (m_0^3 + m_1^3)) (- i+(l + m)\nu )^2
-3\nu (1 - \nu)  \bigl(1 +10 m^{(0,1)}\bigr)\\
\phantom{\Theta_4(l,m|\al)=}{}\times
( -i+2l \nu)(- i+2m \nu)
- 20i \nu (1 + \nu) (2 \nu-1 ) (-i +  (l + m)\nu) m^{(0,1)}\\
\phantom{\Theta_4(l,m|\al)=}{}
 -i\nu  (-i + (l + m) \nu) \bigl(4 \nu^2 +
     5 ( 1 + 24 (m_0^3 + m_1^3)) (1 - \nu)\bigr)\bigr) \al(2-\al)\nn
\\
\phantom{\Theta_4(l,m|\al)=}{}
-16i\nu^4(1-\nu)(2-\nu)\bigl(1+10 m^{(0,1)}\bigr)(l-m) \al(1-\al)(2-\al)+8i(1-\nu)^3\nn
\\
\phantom{\Theta_4(l,m|\al)=}{}\times
(7+120(m_0^3+m_1^3))(-i+(l+m)\nu)(-i+2(l+m)\nu)(-3i+2(l+m)\nu)\nn\\
\phantom{\Theta_4(l,m|\al)=}{}
+8(1-\nu)^2(1+\nu)(2\nu-1)\bigl(1+10 m^{(0,1)}\bigr)\bigl(2(-i+(l+m)\nu)(-i+2(l+m)\nu)\nn
\\
\phantom{\Theta_4(l,m|\al)=}{}
+(-i+2l \nu)(-i+2m \nu)\bigr)
\Bigr),\nn
\end{gather*}
where $m^{(0,1)}=m_0^3+m_1^3+6(m_0+m_1)^2-m_0-m_1$. Unfortunately, we cannot show all the
orders in the expansion of $\Theta$ and other functions from the previous Appendices
that we needed in order to get the formulae (\ref{resM817})--(\ref{resM8det}) because the expressions
become very cumbersome with growing order. Again the ground state result can be obtained
by taking $m_0=m_1=0$.

\subsection*{Acknowledgements}

Our special thanks go to M.~Jimbo, T.~Miwa and F.~Smirnov with whom the work on this paper
was started. Also we would like to thank F.~G{\"o}hmann, A.~Kl{\"u}mper and
S.~Lukyanov for many stimulating discussions. We are grateful to the Volkswagen Foundation
for f\/inancial support.

\pdfbookmark[1]{References}{ref}
\LastPageEnding

\end{document}